\documentclass[10pt,aps,pra,twocolumn,export,superscriptaddress]{revtex4-2} 
\usepackage[colorlinks=true,linkcolor=blue,citecolor=blue,urlcolor=blue,pdftitle={Stitching SIMPS}]{hyperref}

\usepackage[utf8]{inputenc}
\usepackage[german, english]{babel}
\usepackage[dvipsnames]{xcolor}
\usepackage{amsthm}
\usepackage{mathtools}
\usepackage{anyfontsize}
\usepackage{tikz}
\usetikzlibrary{decorations.markings} 
\usetikzlibrary{decorations.shapes,shapes.geometric} 

\usepackage{amsfonts}
\usepackage{amssymb}
\usepackage{braket}
\usepackage{subfigure}
\usepackage[export]{adjustbox} 
\usepackage[inline]{enumitem}  
\usepackage{times}
\usepackage[subscriptcorrection,smallerops]{newtxmath}
\usepackage{microtype}
\usepackage{comment}
\usepackage[normalem]{ulem}

\DeclareMathAlphabet{\mathcal}{OMS}{cmsy}{m}{n}
\DeclareMathAlphabet{\mathsf}{OT1}{cmss}{m}{n}

\renewcommand{\vec}[1]{{\mathbf{#1}}}
\newcommand{\id}{I}

\DeclareMathOperator{\Tr}{Tr}

\DeclareMathOperator{\Span}{span}
\DeclareMathOperator{\spec}{spec}

\newcommand{\CZ}{\mathsf{CZ}}
\newcommand{\CX}{\mathsf{CX}}

\def\Z#1{\mathbb{Z}_{#1}}

\definecolor{opcolor}{RGB}{230, 230, 230}
\definecolor{mpsblue}{RGB}{209, 227, 237}
\definecolor{simpsyl}{RGB}{251, 229, 214}

\newcommand{\ketbra}[2]{|{#1}\rangle 
\langle{#2}|}
\newcommand{\GHZ}{|\text{GHZ}\rangle}
\newcommand{\MG}{|\text{MG}\rangle}
\newcommand{\mpstensor}[3]{%
\begin{tikzpicture}[baseline={(0,-0.1)}, x=1pt, y=1pt]%
        \draw [thick] (-15, 0) -- (15, 0);%
        \draw [thick] (0, 0) -- (0, 15);%
        \draw [thick, fill=mpsblue, rounded corners] (-7, -7) rectangle (7, 7);%
        \node [left] (leftleg) at (-15, 0) {$\strut#2$};%
        \node [right, baseline=(leftleg.base)] at (15, 0) {$\strut#3$};%
        \node [above] at (0, 15) {$#1$};%
    \end{tikzpicture}%
}

\DeclarePairedDelimiter\abs{\lvert}{\rvert}%
\DeclarePairedDelimiter\norm{\lVert}{\rVert}%

\makeatletter
\let\oldabs\abs
\def\abs{\@ifstar{\oldabs}{\oldabs*}}
\let\oldnorm\norm
\def\norm{\@ifstar{\oldnorm}{\oldnorm*}}
\makeatother

\usepackage{bbold}

\graphicspath{{img/}}

\begin{document}

\title{Preparing matrix product states via fusion: constraints and extensions}

\date{\today}

\author{David T. Stephen}
\thanks{These authors contributed equally to this work}
\affiliation{Department of Physics and Center for Theory of Quantum Matter, University of Colorado Boulder, Boulder, Colorado 80309 USA}
\affiliation{Department of Physics and Institute for Quantum Information and Matter, California Institute of Technology, Pasadena, California, USA}

\author{Oliver Hart}
\thanks{These authors contributed equally to this work}
\affiliation{Department of Physics and Center for Theory of Quantum Matter, University of Colorado Boulder, Boulder, Colorado 80309 USA}

\begin{abstract}
    In the era of noisy, intermediate-scale quantum (NISQ) devices, the efficient preparation of many-body resource states is a task of paramount importance. In this paper we focus on the deterministic preparation of matrix-product states (MPS) in constant depth by utilizing measurements and classical communication to fuse smaller states into larger ones. We place strong constraints on the MPS that can be prepared using this method, which we refer to as MPS fusion. Namely, we establish that it is necessary for the MPS to have a flat entanglement spectrum. Using the recently introduced split-index MPS (SIMPS) representation, we then introduce a family of states that belong to interesting phases of matter protected by non-onsite symmetries, {including anomalous and non-invertible symmetries}, and also serve as resources for long-range quantum teleportation, but which lie beyond the scope of ordinary MPS fusion. It is shown constructively that these states can be prepared in constant depth using a broader class of measurement-assisted protocols, which we dub SIMPS fusion. Even in cases when MPS fusion is possible, using SIMPS fusion can give rise to significantly reduced resource overhead. {We also discuss constraints on SIMPS fusion and propose a general framework for fusion that encompasses the MPS and SIMPS protocols.} Our results therefore simultaneously establish the boundaries of conventional MPS fusion and push the envelope of which states can be prepared using measurement-assisted protocols.
\end{abstract}

\maketitle


The complexity of preparing a many-body quantum state from an initial product state is a problem of both fundamental and practical significance. On one hand, the question of whether a many-body state can be prepared using a finite-depth quantum circuit (FDQC) underpins the modern classification of ground-state phases of matter~\cite{Chen2010}. On the other hand, many-body states are essential in many schemes of quantum metrology~\cite{Pezze2018}, communication~\cite{Briegel1998,Popp2005,Acin2007}, and computation~\cite{RaussendorfOneWay,Wei2018}, so any protocol that facilitates their preparation will have important practical consequences.
For these reasons, there has been significant recent interest in developing new protocols to prepare interesting many-body quantum states in quantum computers~\cite{Piroli2021,tantivasadakarn2022longrange,verresen2022efficiently,Tantivasadakarn2023shortest,Lu2022Measurement,iqbal2023topological,bravyi2022adaptive,fossfeig2023experimental,Herringer2023classificationof,gunn2023phases,hart2024playing,piroli2024approximating,Malz2024}. This interest is also driven by the new tools provided by modern quantum computers that go beyond local unitary circuits. One particularly powerful tool is the ability to perform projective measurements on single degrees of freedom and apply unitary operations based on the measurement outcomes~\cite{pino2021demonstration,RyanAndersonQEC2021,Corcoles2021Dynamic,fossfeig2023experimental}. This ability can drastically reduce the circuit depth needed to prepare and manipulate topologically ordered states~\cite{RaussendorfLongRange,tantivasadakarn2022longrange,Lu2022Measurement,bravyi2022adaptive}, allow implementation of long-range quantum gates~\cite{bäumer2023efficient}, and more generally enable exchange between spatial and temporal quantum resources \cite{stephen2022universal,DeCross2023,Hoke2023}.

A powerful method of using measurements to prepare quantum states is based on the matrix product state (MPS) formalism \cite{PerezGarcia2006,Cirac2021}. In Ref.~\cite{Smith_AKLT}, it was shown that the Affleck-Kennedy-Lieb-Tasaki (AKLT) state \cite{Affleck1987}, which is simultaneously a paradigmatic example of an MPS, a symmetry-protected topological (SPT) phase of matter~\cite{Pollmann2012}, and a resource for quantum computation \cite{Brennen2008,Wei2012}, can be deterministically prepared in constant depth using measurement and feedforward. {That is, the same state is prepared in every shot up to a phase, circumventing any need for postselection}. The approach involves preparing small AKLT chains and fusing them into longer chains using measurements. The special symmetry properties of the AKLT state then allow the effects of undesirable measurement outcomes to be undone using unitary feedforward, resulting in a deterministic protocol. Similar constant-depth preparation schemes based on MPS were also given in Refs.~\cite{Piroli2021,gunn2023phases,Malz2024}.
Given the fact that MPS are capable of capturing many interesting classes of one-dimensional (1D) many-body states \cite{Cirac2021}, it is important to understand the limitations and potential extensions of this method, which we refer to as MPS fusion.

In this paper, we derive fundamental constraints on MPS fusion, and then describe a more general method that goes beyond these constraints. Our constraints show that MPS can only be prepared via fusion if they have a flat entanglement spectrum, as is true for the AKLT state. This constraint matches well with the intuition that the fusability of the AKLT state is related to its non-trivial SPT order \cite{Smith_AKLT}, which is in turn associated with entanglement-spectrum degeneracy \cite{Pollmann2010}. For symmetry-breaking MPS with long-range correlations, like the Greenberger-Horne-Zeilinger (GHZ) state, we require that a certain choice of boundary conditions results in a flat entanglement spectrum. 

We then define a family of MPS that do not have a flat entanglement spectrum and therefore cannot be prepared via standard fusion according to our constraints. Nevertheless, we present a generalized approach to fusion that can prepare these states in constant depth. This new approach is based on the recently developed split-index MPS (SIMPS) formalism~\cite{SIMPS}, which was developed to better capture certain features -- including symmetry properties -- of a subset of MPS. Indeed, the class of states for which we construct SIMPS fusion protocols belong to non-trivial phases of matter possessing symmetries that are not a product of single-site unitaries (i.e., non-onsite symmetries), which SIMPS are particularly suited to capturing. We argue that the non-onsite symmetries are the reason why MPS fusion is not possible for these states, thereby demonstrating that SIMPS fusion unlocks the efficient preparation of many-body states belonging to new phases of matter. {We also show how to prepare a large class of states with anomalous symmetries and derive some constraints on states preparable by SIMPS fusion.}

{We finish by showing how MPS fusion and SIMPS fusion can be viewed as instances of a more general framework for fusion. From this perspective, SIMPS fusion goes beyond MPS fusion by
\begin{enumerate*}[label=(\roman*)]
    \item requiring specific boundary conditions,
    \item having more than one round of measurement, and
    \item using unitary feedfoward beyond single-site operators
\end{enumerate*}
}
Finally, we show that there are cases where both protocols work, but the SIMPS fusion protocol requires arbitrarily fewer ancilla qubits to prepare the same state.

The paper is structured as follows. We begin in Sec.~\ref{sec:mps_fusion} by discussing the simplest deterministic, measurement-assisted fusion procedure for MPS both with and without long-range entanglement, using the AKLT and GHZ states as examples, respectively. We then derive constraints on when MPS fusion can succeed in a more general setting, in particular allowing for generalized two-qudit fusion measurements and more general kinds of unitary feedforward. Given these constraints, in Sec.~\ref{sec:non_fusible} we highlight two states with appealing physical properties that cannot be prepared using standard MPS fusion. We then introduce the SIMPS formalism in Sec.~\ref{sec:SIMPS} and show that the two examples from Sec.~\ref{sec:non_fusible} can be represented more efficiently as SIMPS. In Sec.~\ref{sec:simps_fusion} we show that the two example states \emph{can} be fused deterministically as SIMPS by employing a two-step fusion procedure involving feedforward corrections that have the form of FDQCs.
{The generalization of this protocol to a more general family of states that are
invariant under anomalous symmetries is presented in Sec.~\ref{sec:omega_simps}, and constraints on which states can be prepared by SIMPS fusion are outlined in Sec.~\ref{sec:simps_constraints}.
Finally, we present a generalized framework for fusion of 1D states and}
compare the resource requirements for MPS and SIMPS fusion in Secs.~\ref{sec:unifying_framework} and \ref{sec:comparison}, respectively. We close with a discussion of our results and their implications.


\section{MPS Fusion}
\label{sec:mps_fusion}

We begin by reviewing how two MPS satisfying certain criteria can be deterministically fused using measurements and unitary feedforward~\cite{Smith_AKLT}. {We refer to this specific approach to state preparation as ``MPS fusion.''} Combining many such fusions in parallel allows for arbitrarily large states to be prepared deterministically and exactly using constant-depth quantum circuits, sidestepping the bounds that constrain unitary preparation protocols~\cite{Bravyi2006,Chen2010}. A translation-invariant MPS for $N$ qudits on periodic boundary conditions (PBC) is defined as
\begin{equation}
    |\psi\rangle = \sum_{i_1,\dots,i_N} \mathrm{Tr}(A^{i_1}A^{i_2}\cdots A^{i_N})|i_1,i_2,\dots,i_N\rangle \, ,
\end{equation}
where $A^i$ is a $D \times D$ matrix acting on the \emph{virtual space}, and each physical index runs over $i_k = 0, 1, \dots, d-1$. The parameter $D$ is called the MPS bond dimension. Throughout this paper, we will mainly be concerned with MPS with open boundary conditions (OBC), which we define as
\begin{equation}
    \ket{\bar{\psi}} = 
    \sum_{\substack{i_1, \dots, i_N \\ a,b}} 
    \braket{a | A^{i_1} A^{i_2} \cdots A^{i_{N}} | b} \ket{a, i_1, i_2, \dots, i_N, b}
    \, ,
    \label{eqn:MPS}
\end{equation}
where we have terminated the chain at either end with a $D$-dimensional qudit whose states are labeled by $a$ and $b$. We refer to the $d$- and $D$-dimensional qubits as the bulk and boundary qudits, respectively. 

The protocols we discuss are best illustrated using the graphical representation of MPS. The three-index tensor $A$ is represented graphically as
\begin{equation} \label{eqn:MPS-tensor}
    \mpstensor{i}{a\!}{\!b} = A_{ab}^i .
\end{equation}
The upper leg represents the physical index, and the left and right legs are the virtual indices. Then, the MPS on OBC \eqref{eqn:MPS} is represented for $N=4$, as
\begin{equation} \label{eqn:MPS-fig}
   \includegraphics[scale=0.2,raise=-1.3ex]{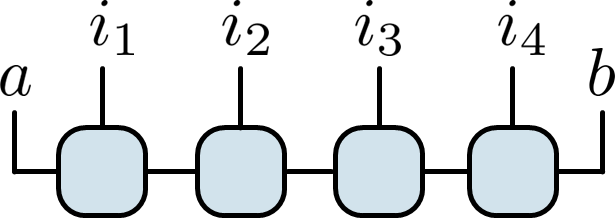}\ ,
\end{equation}
where we have bent the virtual legs at the boundary to emphasize that they correspond to physical degrees of freedom. For notational simplicity, we will often omit the index labels.


\subsection{AKLT state}
\label{sub:AKLT-prep}

To illustrate the general features of the MPS fusion procedure, we work with the AKLT~\cite{AKLT1987,AKLT1988} state, studied in detail in Ref.~\cite{Smith_AKLT}.
The AKLT state is defined by the following $D=2$ matrices for a particular physical basis of the $d=3$ qudits,
\begin{equation}
    A^{0} = \frac{1}{\sqrt{3}} X , \quad 
    A^{1} = \frac{1}{\sqrt{3}} Y , \quad 
    A^{2} = \frac{1}{\sqrt{3}} Z . 
    \label{eqn:AKLT-MPS}
\end{equation}
The MPS~\eqref{eqn:AKLT-MPS} is a paradigmatic example of a state possessing SPT order~\cite{Pollmann2012} and hence cannot be prepared (even approximately) by a symmetry-respecting FDQC~\cite{Chen2010,Huang2015}.
The $\Z2 \times \Z2$ symmetry of the AKLT state, which with PBC is generated by, e.g., rotation by $\pi$ about the $x$ and $z$ axes, $\prod_{k} \exp(i\pi S^x_k)$ and $\prod_{k}\exp(i\pi S^z_k)$, manifests as a local symmetry of the tensor~\eqref{eqn:AKLT-MPS},
\begin{equation}
    \mpstensor{U_P}{P\!}{\!P}
    \!,
    \label{eqn:AKLT-symm}
\end{equation}
up to a global phase, where $P \in \{I, X, Y, Z\}$ are the Pauli operators and, e.g., $U_X = \exp(i\pi S^x)$~\cite{PerezGarcia2008,Pollmann2010}. Thus, application of a unitary operator to the physical degrees of freedom can be converted via~\eqref{eqn:AKLT-symm} to Pauli operators acting in the virtual space. Equivalently, since $P^2 = I$, a Pauli operator acting on the right virtual leg can be pushed to the left leg (and vice versa) by the same mechanism. 

\begin{figure}
    \centering
    \includegraphics[width=0.6\linewidth]{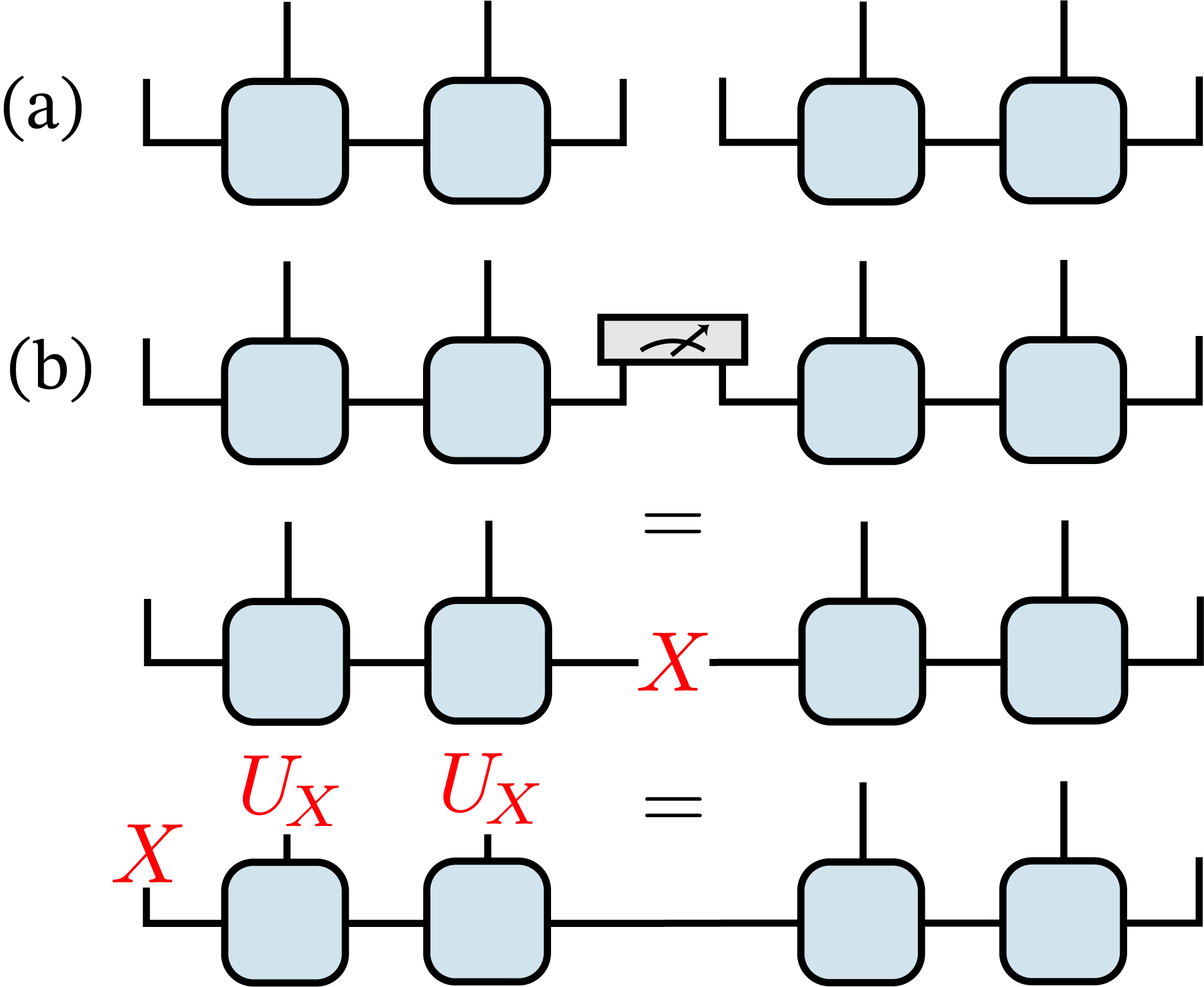}
    \caption{Fusing two copies of the AKLT state. Undesirable measurement outcomes result in the red Pauli operators, which are corrected using the symmetries~\eqref{eqn:AKLT-symm} of the MPS tensor. (a) Initial state: two short, disconnected AKLT chains. (b) Fuse the two intervening boundary legs using a Bell measurement. Depending on the measurement outcome, a Pauli operator $P$ is inserted on the fused leg, shown for the case $P=X$. This erroneous operator can be pushed to the boundary using onsite $U_P$ operators [Eq.~\eqref{eqn:AKLT-symm}]. Finally, these operators can be removed by applying the corresponding $U_X$ and boundary $X$ operators.}
    \label{fig:aklt_fusion}
\end{figure}

Two preexisting, disjoint AKLT states can be fused deterministically by employing a Bell measurement between the boundary qudits of the two chains and feedforward of the measurement outcomes~\cite{Smith_AKLT}.
As depicted in Fig.~\ref{fig:aklt_fusion}, this fuses the two boundary legs into a single virtual leg, up to a Pauli defect, the presence of which is heralded by the measurement outcome. Using the symmetries~\eqref{eqn:AKLT-symm} of the tensor $A$, such defects can be paired up or pushed to the boundaries by application of $U_P$ to the physical degrees of freedom. The initial AKLT states of size $N=O(1)$ can be prepared, e.g., using a sequential unitary scheme~\cite{Schon2005,Smith_AKLT}.

To generate an AKLT state on PBC, we perform the same fusion measurement on the final pair of virtual legs to close the boundaries. However, this means we no longer have any open virtual legs. 
As a result, there is nowhere to push byproducts in order to remove them.
Instead, we can use the same pushing-through operation to cancel byproducts pairwise, leaving behind at most one byproduct. The resulting MPS with one Pauli $P$ sitting somewhere in the virtual space is a symmetry-twisted version of the AKLT state~\cite{Kapustin2017}. If we want the untwisted version, we need $P=I$, which can be achieved with postselection. Importantly, the amount of postselection is independent of system size, since we can keep $1/4$ of experiments for all $N$.

A simple argument showing that the AKLT state cannot be made \emph{exactly} with an FDQC comes from the fact that it has exponentially decaying correlations. This implies a nonzero correlation length, whereas any state generated by an FDQC has strictly zero correlations outside the circuit Lieb-Robinson light cone~\cite{AnthonyLR2023}. This demonstrates the interesting fact that, by incorporating measurements, we can construct exact states with nonzero correlation length in constant depth. Note that this argument applies even to circuits that do not respect the symmetry, but does not prohibit approximate preparation with an FDQC employing symmetry-breaking gates.

\subsection{GHZ state}
\label{sub:GHZ-prep}

Next, we illustrate the MPS-fusion approach to preparing the GHZ state. This is not the optimal way to prepare a GHZ state in terms of resource overhead, as discussed in Sec.~\ref{sec:comparison}, but it nonetheless serves as another example where MPS fusion is possible. The state is defined as,
\begin{equation}
    |\text{GHZ}\rangle = \frac{1}{\sqrt{2}}(\ket{00\cdots 0} + \ket{11\cdots 1})\ ,
\end{equation}
and involves a macroscopic superposition of MPS and, correspondingly, exhibits long-range correlations and entanglement. The MPS representation of the GHZ state
is constructed from the rank-1 matrices
\begin{equation}
    A^0 = |0\rangle \langle 0|, \quad A^1 = |1\rangle \langle 1|
    \, .
\end{equation}
The GHZ MPS tensor obeys the following symmetry relations:
\begin{equation}
    \mpstensor{X}{X\!\!}{\!\!X}
    ,
    \qquad 
    \mpstensor{I}{Z\!\!}{\!\!Z}
    .
    \label{eqn:GHZ-symm}
\end{equation}
The first encodes the global symmetry $\prod_i X_i$ of the state, while the second is physically associated with the long-range order of the state. 
Since $X$ and $Z$ generate all Pauli operators, the symmetries~\eqref{eqn:GHZ-symm} can be used to push any Pauli. In contrast to the AKLT example, the symmetries imply that $Z$ operators in the virtual space can be pushed ``for free,'' i.e., without application of any unitary operator to the physical legs. Hence, as in Sec.~\ref{sub:AKLT-prep}, given $N$ copies of the tensor $A^i_{ab}$ interpreted as a three-qubit state $\ket{A}$, the $N$-qubit GHZ state (plus two boundary qubits) can be prepared deterministically by 
\begin{enumerate*}[label={(\roman*)}]
    \item fusing the interstitial legs using a Bell measurement,
    \item pairing up or pushing to the edge any $P \neq I$ using~\eqref{eqn:GHZ-symm}.
\end{enumerate*}
The initial states $\ket{A}$ are simply three-qubit GHZ states and can be prepared in parallel unitarily.
As discussed in Ref.~\cite{Smith_AKLT}, the disentangled qubits resulting from Bell measurements can be reincorporated into the $N$-qubit GHZ state using reset operations and entangling gates.


\subsection{Constraints on MPS fusion}
\label{sec:mps_fusion_constraints}

Here, we derive the constraints on when MPS fusion is possible. First, we define the class of fusion protocols that we will consider. Our goal is to prepare the OBC MPS \eqref{eqn:MPS} defined by a tensor $A^i$. We won't discuss the PBC case, but we note that it can be obtained from the OBC MPS with some $O(1)$ postselection overhead, as in the AKLT example (i.e., by postselecting on the outcome of the final fusion measurement). The most general MPS that we will try to prepare by fusion is defined by the matrices,
\begin{equation} \label{eq:mps_tensor}
    A^i=\bigoplus_{\alpha=1}^{N_b} A^i_\alpha\, ,
\end{equation}
where $A^i_\alpha$ are $D_\alpha\times D_\alpha$ matrices (with $\sum_\alpha D_\alpha = D)$ and the $A^i$ are block injective, see Appendix \ref{sec:ent_spec}.
This implies that the MPS generated by $A^i$ can be written as $|\psi\rangle = \sum_\alpha |\psi_\alpha\rangle$ where each $|\psi_\alpha\rangle$ is short-range entangled and $\lim_{N\to\infty}\langle \psi_\alpha|\psi_\beta\rangle = 0$  when $\alpha\neq\beta$. We say an MPS is \textit{normal} if $N_b=1$, and otherwise it is \textit{non-normal}~\cite{Cirac2021}.
Physically, normal MPS (like the AKLT state) have a finite correlation length, whereas non-normal MPS (like the GHZ state) have long-range correlations and long-range entanglement. Interestingly, the protocols for fusing these two distinct kinds of MPS are largely the same, as seen in the two examples shown in Sec.~\ref{sec:mps_fusion}. 

A general MPS fusion protocol proceeds as follows. Starting with a product of many three-leg MPS tensors
of the form Eq.~\eqref{eq:mps_tensor},
we apply fusion measurements to pairs of neighbouring virtual legs.
The most general fusion measurements we consider will consist of applying a two-site unitary to the two virtual legs to be fused, followed by projective measurements of both legs in the computational basis. This applies the following fusion operation to the pair of legs,
\begin{equation}
    \begin{tikzpicture}[baseline={(0,0)}, x=1pt, y=1pt]
        \draw [thick] (-15, -20) -- (-15, 20);
        \draw [thick] (15, -20) -- (15, 20);
        \node (rect) at (0, 0) [draw,thick,minimum width=40pt,minimum height=15pt,fill=opcolor,rounded corners] {$U$};
        \node [above] at (-15, 20) {$\bra{s_1}$};
        \node [above] at (15, 20) {$\bra{s_2}$};
    \end{tikzpicture}
    \equiv
    \frac{1}{\sqrt{D}}
    \:\:
    \begin{tikzpicture}[baseline={(0,0)},x=1pt, y=1pt]
        \draw [thick] (-20, -20) -- (-20, 0) -- (20, 0) -- (20, -20);
        \node (rect) at (0, 0) [draw,thick,minimum width=20pt,minimum height=15pt,fill=opcolor,rounded corners] {$V^{\vec{s}}$};
    \end{tikzpicture}
    \label{eqn:2-qubit-measurement}
\end{equation}
The operators $V^\vec{s}$ with $\vec{s} \equiv (s_1, s_2)$ are the byproduct operators that need to be pushed through the virtual space. For a Bell measurement, as in the AKLT protocol, these are the Pauli operators. In a more general case, we see from Eq.~\eqref{eqn:2-qubit-measurement} that the set of byproduct operators must always satisfy completeness $\sum_\vec{s} V^\vec{s}_{ij} \bar{V}^\vec{s}_{kl} = D \delta_{ik}\delta_{jl}$ and orthonormality $\Tr( V^{\vec{s}\dagger} V^{\vec{s}'})  = D\delta_{\vec{s}\vec{s}'}$. Since $V^\vec{s}$ are orthonormal matrices, they form a basis for the space of all $D \times D$ matrices.

After the fusion measurement, the next step we allow in the protocol involves applying unitaries to the bulk qudits to push the byproducts to the boundary qudits, and then undoing the byproducts on the boundary. Since we want to undo the byproduct operators at the end, they should be unitary themselves~\footnote{{We remark that non-unitary operators can be applied to the boundaries using postselection. See, e.g., Ref.~\cite{Lin2021}.}}.
We also assume for now that the byproduct operators are dealt with using onsite push-through relations like the one in Eq.~\eqref{eqn:AKLT-symm}, which generalizes to,
\begin{equation} \label{eq:mps_symm}
    \sum_j u^i_jA^j = VA^iV^{\dagger} \, ,
\end{equation}
for some unitaries $u$ and $V$. 
Using this equation, we can apply $u$ to the physical leg of the MPS tensor to push $V$ from one virtual leg to the other. 
In Appendix~\ref{app:asymmetric-pushthrough} we show that identical constraints can be derived when
Eq.~\eqref{eq:mps_symm} is relaxed to asymmetric push-through relations ($V$ may be pushed to a different unitary $W$) and to cases where the unitary operator used to push is not translationally invariant and not a product of onsite operators.

Our goal is to constrain MPS that have enough push-through relations of the form~\eqref{eq:mps_symm} to remove all possible byproduct operators resulting from the fusion measurements. The starting point for the derivation of our constraints is the MPS transfer matrix $T$, defined as,
\begin{equation} \label{eq:tm}
    T = \sum_i A^i\otimes \bar{A}^i \, .
\end{equation}
If Eq.~\eqref{eq:mps_symm} holds for some $u$ and $V$, then $(V\otimes \bar{V})T(V^\dagger \otimes V^T) = T$. This implies that $V\otimes \bar{V}$ maps fixed-points of $T$ to fixed-points such that,
\begin{equation} \label{eq:v_constraint}
    (V\otimes \bar{V})\Pi(V^\dagger \otimes V^T) = \Pi
    \, ,
\end{equation}
where {$\Pi=\sum_\alpha\ketbra{R_\alpha}{L_\alpha}$ is the projector onto the space spanned by the fixed-points of $T$, where the left and right fixed-points, $\bra{L_\alpha}$ and $\ket{R_\alpha}$, respectively, are normalized such that $\braket{L_\alpha | R_\beta} = \delta_{\alpha\beta}$}. Taking the partial trace of Eq.~\eqref{eq:v_constraint} over the second subsystem in the tensor product gives,
\begin{equation} \label{eq:partial_projector}
    V \tilde{\Pi} V^{\dagger} = \tilde{\Pi}
    \, ,
\end{equation}
where 
$\tilde{\Pi}=\mathrm{Tr}_2\Pi$
is the partial trace of $\Pi$.

Now, if we have a push-through relation \eqref{eq:mps_symm} for $V=V^\vec{s}$ for all measurement outcomes $\vec{s}$, then Eq.~\eqref{eq:partial_projector} implies that $\tilde{\Pi}$ commutes with all $V^\vec{s}$~\footnote{Note that our assumption of block injectivity combined with the existence of the right-canonical form \cite{PerezGarcia2006} implies that every measurement outcome occurs with non-zero probability for sufficiently large systems.}. Since the $V^\vec{s}$ span the space of all $D\times D$ matrices, this is only possible if $\tilde{\Pi}$ is proportional to the identity. 
Physically, this constraint has implications for the entanglement spectrum of fusible MPS.
Note that we can always write the right and left fixed-points of $T$ in terms of matrices $\sigma_{R_\alpha}$ and $\sigma_{L_\alpha}$, i.e., defining
$|X\rangle = \sqrt{D}(\sigma_{X}\otimes I)|\Omega\rangle$, with $\ket{\Omega}$ the $D$-dimensional qudit Bell state with normalization $\braket{\Omega | \Omega} = 1$.
Then, we can write,
\begin{equation}
    \tilde{\Pi} = \bigoplus_{\alpha=1}^{N_b} \sigma_{R_\alpha}\sigma_{L_\alpha}^T\, .
\end{equation}
In Appendix~\ref{sec:ent_spec}, we show that the spectrum of $\tilde{\Pi}$ is exactly the entanglement spectrum of the OBC MPS in the thermodynamic limit. 
Therefore, the constraint $\tilde{\Pi}\propto I$ means that \textit{any fusible MPS must have a flat entanglement spectrum}. This is our main result that we will use to constrain MPS fusion.

In the case of non-normal MPS fusion, the constraint $\tilde{\Pi} \propto I$ has further consequences that give important intuition. From Eq.~\eqref{eq:v_constraint} we deduce that the Bell states $|\Omega_d\rangle = d^{-1/2}\sum_{i=0}^{d-1} |ii\rangle$ and $\langle \Omega_d|$ must be right and left fixed-points, respectively, for each diagonal block of the MPS (see Appendix~\ref{app:flat-spectrum} for details). Normalization of the fixed-points then implies 
\begin{equation} \label{eq:equal_size_constraint}
    \tilde{\Pi} = \bigoplus_{\alpha=1}^{N_b} D_\alpha^{-1} I_\alpha
    \, ,
\end{equation}
with $I_\alpha$ the $D_\alpha \times D_\alpha$ identity matrix.
Hence, \emph{it is only possible to satisfy $\tilde{\Pi} \propto I$ if the blocks being fused are of equal size}, i.e., $D_\alpha = D/N_b$ for all $\alpha$.
Given the close connection between bond dimension and entanglement, this means that MPS fusion can only prepare non-normal MPS that are superpositions of short-range-entangled states, where each state has roughly the same amount of entanglement. This is true for GHZ states, where each state in the superposition is a product state, but will not be true for an example we present later.

We stress that the above constraints are only \emph{necessary} conditions for fusibility; we do not attempt to provide also sufficient conditions here. For instance, we do not ask that the different correction unitaries $u$ commute with each other, which in general means that the correction step could require a circuit depth that is linear in the number of fusion measurements.

In the rest of the paper, we show how a more general notion of fusion can be used to get around the above constraints. We remark that MPS with non-degenerate entanglement spectrum were also generated with measurements in constant depth using a different method in Ref.~\cite{Zhu2023}.

\section{Examples of states beyond MPS fusion}
\label{sec:non_fusible}

Having understood how the MPS fusion protocol is constrained, we now consider a pair of examples that {cannot be prepared via MPS fusion}. These states are interesting for two reasons. First, they belong to phases of matter protected by exotic symmetries which are not products on onsite unitaries, which relates interesting physics involving anomalies and dualities. Second, they can also serve as resources for long-range quantum teleportation and measurement-based quantum computation like the AKLT state.

\subsection{Non-normal example}

Our first example is a non-normal MPS. We consider the following state defined on PBC,
\begin{equation}
    |\phi\rangle = \frac{1}{\sqrt{\mathcal{N}}}(\ket{++\cdots +} + |C\rangle)
    \label{eqn:non-normal-state}
\end{equation}
for some $N$-dependent normalization factor $\mathcal{N}$, where $|C\rangle$ is the 1D cluster state \cite{Raussendorf2001} defined as,
\begin{equation}
    |C\rangle = U_{\CZ} \ket{++\cdots +}
    \, ,
    \label{eqn:cluster-state}
\end{equation}
where $\ket{+}=(\ket{0}+\ket{1})/\sqrt{2}$,  $U_{\CZ}=\prod_{i=1}^N \CZ_{i,i+1}$, and $\CZ= |0\rangle\langle 0|\otimes I + |1\rangle\langle 1|\otimes Z$ is the controlled-$Z$ operator. Since $|\phi\rangle$ is a macroscopic superposition of MPS, it has long-range correlations and is non-normal. 
Because the two states in the superposition are exchanged by $U_{\CZ}$, the state $\ket{\phi}$ is associated with the spontaneous breaking of this symmetry, which is non-onsite in the sense that it is not a product of single-site unitaries, even after blocking. Such non-onsite symmetries are often associated with anomalies on the lattice \cite{Chen2011,Cheng2023}, {and we strengthen the connection to anomalies in Sec.~\ref{sec:omega_simps}}. We claim that \textit{any} state with this property cannot be prepared with MPS fusion. Roughly speaking, any $U_{\CZ}$ symmetry-breaking state can be written as $|\psi\rangle + U_{\CZ}|\psi\rangle$ for some state $|\psi\rangle$. The state $U_{\CZ}|\psi\rangle$ will generically be more entangled than the state $|\psi\rangle$, meaning it will require a larger MPS bond dimension. Therefore, the non-normal MPS which describes the superposition of the two states will have blocks of different sizes, which we have argued above is incompatible with MPS fusion. We will make argument this more precise for the state $|\phi\rangle$ shortly.

The state $|\phi\rangle$ is also closely related to other states of interest. For instance, applying a local change of basis maps the state to another important example of an MPS: the Majumdar-Ghosh state \cite{Majumdar1969} (see Appendix~\ref{sec:mg}). Also, by applying the so-called Kennedy-Tasaki transformation \cite{Kennedy1992,Else2013,mana2024kennedytasaki}, we can map the state~\eqref{eqn:non-normal-state} to $(\ket{++\cdots +} + \GHZ)^{\otimes 2}$. This state is interesting since it is associated with the spontaneous breaking of a non-invertible symmetry, namely the Kramers-Wannier duality, which interchanges $\ket{++\cdots +} \leftrightarrow \GHZ$ \cite{seiberg2024noninvertible}. Since the Kennedy-Tasaki transformation can be implemented by an FDQC and the Kramers-Wannier transformation \cite{mana2024kennedytasaki}, and the latter can be implemented in constant depth with measurements and feedforward \cite{tantivasadakarn2022longrange}, our protocol for preparing $|\phi\rangle$ also gives a constant-depth protocol for preparing (two copies of) $\ket{++\cdots +} + \GHZ$.

We can write $\ket{\phi}$ as an MPS by taking a direct sum of the MPS matrices for the product state $\ket{++\cdots+}$ and $|C\rangle$ \cite{Cirac2021},
\begin{equation} \label{eq:non_normal_mps}
    A^0 = \frac{1}{\sqrt{2}}
    \begin{pmatrix}
        1 & 0 & 0 \\
        0 & 1 & 1 \\
        0 & 0 & 0
    \end{pmatrix}
    \, , \quad 
    A^1 = \frac{1}{\sqrt{2}}
    \begin{pmatrix}
        1 & 0 & 0 \\
        0 & 0 & 0 \\
        0 & 1 & -1
    \end{pmatrix}
    \, .
\end{equation}
Using standard MPS technology \cite{PerezGarcia2006}, we can find a parent Hamiltonian $H = \sum_i h_i$ for $|\phi\rangle$ where,
\begin{equation}
    h_i = (3-X_i)(1 - Z_{i-1}X_iZ_{i+1}) - 2P_i^+(X_{i-1}X_{i+1}+Y_{i-1}Y_{i+1})
    \label{eqn:non-normal-parent}
\end{equation}
where $P_i^+=(1+X_i)/2$ is the projector onto the state $\ket{+}$ on site $i$. By construction, the Hamiltonian $H$ annihilates both $\ket{++\dots+}$ and the cluster state $\ket{C}$, while all other eigenstates are separated by a gap. The Hamiltonian defined by~\eqref{eqn:non-normal-parent} is maximally local but is not symmetric under the non-onsite symmetry $U_{\CZ}$. A symmetric parent Hamiltonian can be constructed from the symmetrized local terms $h_i + U_{\mathsf{CZ}}h_iU_{\mathsf{CZ}}$. We could also construct a parent Hamiltonian by starting from that of the Majumdar-Ghosh state or of $\ket{++\cdots +} + \GHZ$ \cite{OBrien2018} and applying the transformations discussed above.

As discussed above, the two blocks of the MPS have different sizes (1 vs.~2), so we expect that MPS fusion is not possible. To show this explicitly, we compute the projector onto the fixed-point space of the MPS transfer matrix, 
\begin{equation}
    \Pi = |00\rangle\langle 00| + \frac12(|11\rangle + |22\rangle)(\langle 11| + \langle 22|)
    \, .
\end{equation}
Note that, since the MPS~\eqref{eq:non_normal_mps} is non-normal, the transfer matrix has degenerate fixed-points.
Taking the partial trace over the second subsystem $\tilde{\Pi} = \Tr_2\Pi$, we obtain
\begin{equation} \label{eq:pitilde}
    \tilde{\Pi} = |0\rangle\langle 0| + \frac12 (|1\rangle\langle 1| + |2\rangle\langle 2|) \, .
\end{equation}
Since $\tilde{\Pi}$ is not equal to the identity, the entanglement spectrum of the MPS is not flat, and we see from Sec.~\ref{sec:mps_fusion_constraints} that the state~\eqref{eqn:non-normal-state} cannot be prepared using MPS fusion.

An alternative way to understand the failure of MPS fusion is to constrain the byproduct operators that can be pushed through the virtual space using Eq.~\eqref{eq:partial_projector}. Specifically, commutation of $V$ with $\tilde{\Pi}$ enforces that $V$ is block diagonal with the  block structure
\begin{equation} \label{eqn:V-block-structure}
    V= \begin{pmatrix}
        1& 0& 0 \\
        0&\cdot & \cdot \\
        0 &\cdot & \cdot
    \end{pmatrix}
    \, .
\end{equation}
This is clearly not sufficient to correct for all byproduct operators that would result from a qutrit Bell measurement, which include, e.g., $\mathcal{X} = |0\rangle\langle 1| + |1\rangle\langle 2| + |2\rangle \langle 0|$.
As an example of an operator that is pushable, observe that the tensor~\eqref{eq:non_normal_mps} obeys the symmetry $A^i = W A^i W^\dagger$ where,
\begin{equation} \label{eq:w-matrix}
    W = 
    \begin{pmatrix}
        1 & 0 & 0 \\
        0 & -1 & 0 \\
        0 & 0 & -1
    \end{pmatrix}
    \, .
\end{equation}

\subsection{Normal example}

The second example MPS is defined by the following $D=3$ matrices:
\begin{equation}
    A^0 = \frac{1}{\sqrt{2}}
    \begin{pmatrix}
        1 & 0 & 0 \\
        0 & 0 & 0 \\
        0 & 1 & 1
    \end{pmatrix}
    \, , \quad 
    A^1 = \frac{1}{\sqrt{2}}
    \begin{pmatrix}
        0 & 1 & -1 \\
        -1 & 0 & 0 \\
        0 & 0 & 0
    \end{pmatrix}
    \, .
    \label{eqn:nice-mps}
\end{equation}
A closely related example was presented in Ref.~\cite{Wahl2012}. Like the AKLT state, the state~\eqref{eqn:nice-mps} is a normal MPS with non-zero correlation length, so it cannot be prepared exactly using an FDQC. The state also has computational utility, since it can act as a resource for long-range quantum teleportation and measurement-based quantum computation~\cite{SIMPS}.
As before, we can use the physical properties of the state to argue that it cannot be prepared via MPS fusion. 
While the non-normal MPS \eqref{eqn:non-normal-state} belongs to a symmetry-breaking phase of the non-onsite symmetry $U_{\CZ}$, we will show in Sec.~\ref{sec:SIMPS} that the normal MPS defined by Eq.~\eqref{eqn:nice-mps} has non-trivial SPT order protected by $U_{\CZ}$ and another $\Z2$ symmetry. In Ref.~\cite{SIMPS}, it was shown that SPT order with non-onsite symmetries is not associated with entanglement spectrum degeneracy. Therefore,  the results of Sec.~\ref{sec:mps_fusion_constraints} would imply that MPS fusion is not possible.

To confirm this intuition, we can straightforwardly obtain the unique fixed-points of the transfer matrix corresponding to Eq.~\eqref{eqn:nice-mps},
\begin{subequations} \label{eqn:fixed_pts}
    \begin{align} 
    |L\rangle &= \frac{1}{\sqrt{3}}(|00\rangle + |11\rangle + |22\rangle), \\
    |R\rangle &= \frac{1}{\sqrt{6}}(2|00\rangle + |11\rangle + |22\rangle).
\end{align}
\end{subequations}
Observe that, while $|L\rangle$ is a qutrit Bell state, $|R\rangle$ is not. We can then compute that $\tilde{\Pi}$ has the same form as in Eq.~\eqref{eq:pitilde}. Hence, the entanglement spectrum is not flat, and we again deduce that the state cannot be fused as an MPS. As in the non-normal case, this can also be seen by the fact that pushable byproducts $V$ must have a block diagonal form. In this case, the matrix $W$ in Eq.~\eqref{eq:w-matrix} corresponds to a global onsite $Z$ symmetry since,
\begin{equation}
    \sum_{j} Z_{ij}A^j = W A^i W^\dagger \, .
\end{equation}


\section{SIMPS}
\label{sec:SIMPS}

The two example states presented in the previous section share a common property: they possess the non-onsite symmetry $U_{\CZ}$. Symmetries played an important role in the AKLT and GHZ fusion protocols describes in Sec.~\ref{sec:mps_fusion}, so one might expect that $U_{\CZ}$ could play a role in a fusion protocol. However, while the MPS formalism is able to capture such symmetries \cite{Chen2011,Cirac2021}, the treatment is more complicated than the case of onsite symmetries. To address this, we use the recently introduced SIMPS formalism \cite{SIMPS} which is able to capture such non-onsite symmetries more elegantly than MPS can. In this section, we review SIMPS and their graphical representation and show how to convert an MPS to a SIMPS. We then describe how the SIMPS representation motivates an alternative two-step, measurement-assisted preparation procedure that takes full advantage of the non-onsite symmetries of the states, and demonstrate this procedure for the examples presented in the previous section.

\subsection{Introduction to SIMPS}

With periodic boundaries, a SIMPS defined on $N$ qudits of local Hilbert space dimension $d$ is~\cite{SIMPS},
\begin{equation} \label{eq:pbc_simps}
    \ket{\psi} = \sum_{i_1, \dots, i_N} \Tr (B^{i_1 i_2} B^{i_2 i_3} \cdots B^{i_{N} i_1}) \ket{i_1 i_2 \cdots i_N}
    \, ,
\end{equation}
where $B^{ij}$ is a $\chi^i \times \chi^j$ matrix.
The examples that we consider have uniform $\chi^i = \chi$, referred to as the SIMPS bond dimension.
With open boundaries, 
\begin{equation} \label{eq:obc_simps_eq}
    \ket{\bar{\psi}} = \sum_{\substack{i_1, \dots, i_N \\ a, b}} \braket{a | B^{i_1 i_2} B^{i_2 i_3} \cdots B^{i_{N-1} i_N} | b} \ket{a, i_1 i_2 \cdots i_N, b}
    \, ,
\end{equation}
which includes $\chi$-dimensional qudits at the left and right boundaries of the system indexed by $a$ and $b$ that run from $0$ to $\chi-1$. It is helpful to introduce a graphical description of SIMPS. We denote the four-index tensor as,
\begin{equation}
    \includegraphics[scale=0.35,valign=c]{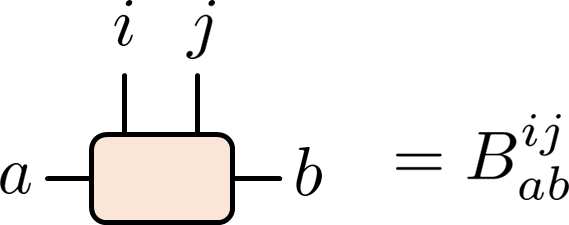}
\end{equation}
We will also make extensive use of $\delta$-tensors, denoted by a small black circle, which enforce that all indices are in the same state in the computational basis $\{|0\rangle,\dots,|d-1\rangle\}$,
\begin{equation} \label{eqn:delta-tensor}
    \includegraphics[scale=0.175,valign=c]{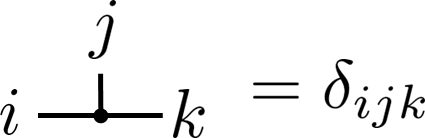}
\end{equation}
Using~\eqref{eqn:delta-tensor}, we can depict a SIMPS with open boundaries (for $N=5$) as,
\begin{equation} \label{eq:obc_simps}
    \includegraphics[scale=0.35,valign=c]{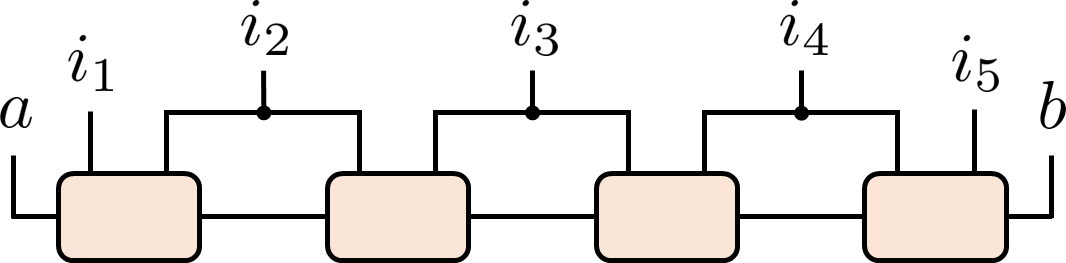}
\end{equation}
Each physical index is split across two adjacent SIMPS tensors $B^{ij}$. We note an important property of $\delta$-tensors that we will use throughout this work. Since all indices connected by a $\delta$-tensor take the same value in the $Z$ basis, any controlled unitary that is controlled by one of the indices can equivalently be controlled by any other index that is connected via a $\delta$ tensor. For example,
\begin{equation} \label{eqn:delta_trick}
    \includegraphics[scale=0.2,valign=c]{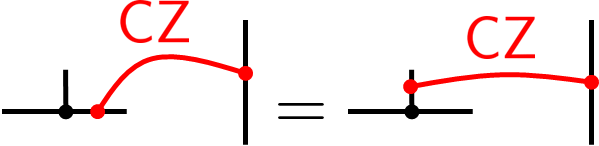}\ .
\end{equation} 

MPS and SIMPS are equivalent in the sense that any state that can efficiently be represented as an MPS can also be efficiently represented as a SIMPS, and vice versa. This can be shown using explicit procedures to convert between MPS and SIMPS \cite{SIMPS}. To convert an MPS to a SIMPS, we first write $A^i=A^iP^iP^{i\dagger}$ where $P^iP^{i\dagger}$ is the projector onto the rowspace of $A^i$, and $P^i:\mathbb{C}^{\chi^i} \rightarrow \mathbb{C}^D$ is an isometry satisfying $P^{i\dagger}P^i=\id_{\chi^i}$ where $\chi^i$ is the rank of $A^i$. 
{Note that the matrix $P^i$ may be systematically obtained from the right singular vectors of the matrix $A^i$.}
Then we define the SIMPS tensor,
\begin{equation} \label{eq:mpstosimps}
    B^{ij}=P^{i\dagger}A^jP^j
    \, .
\end{equation}
The SIMPS generated by the tensor $B^{ij}$ is the same state as the MPS generated by $A^i$ on periodic boundaries. On open boundaries, however, the states will differ near the boundaries, which is the important fact that enables our fusion protocols. The above procedure demonstrates the important fact that \textit{the MPS that are amenable to a SIMPS representation are those for which the matrices $A^i$ are not full rank}. 
Since the rank of $A^i$ depends on the choice of physical basis $\{i\}$, the SIMPS bond dimension also depends on the physical basis. As a consequence, it is generally necessary to identify an ideal physical basis for fusing SIMPS.

In the SIMPS representation, each physical index is ``split'' between two neighboring tensors. One could also imagine splitting indices further, say between three neighboring tensors. However, after sufficient blocking (i.e., increasing the size of the unit cell), one always recovers the form of Eq.~\eqref{eq:pbc_simps} \cite{SIMPS}. Similarly, while we only consider non-onsite symmetries generated by nearest-neighbor quantum circuits, all finite-depth circuits can be made nearest-neighbor after sufficient blocking, so this case is generic.

\subsection{Non-fusible MPS as SIMPS}

Now, we show how to write the two non-fusible MPS from Sec.~\ref{sec:non_fusible} as SIMPS. We then describe the symmetry properties of the SIMPS tensors that we will use for SIMPS fusion. We focus on the normal example and discuss the minor modifications needed to understand the non-normal example at the end.

\begin{figure}[t]
    \centering
    \includegraphics[scale=0.175]{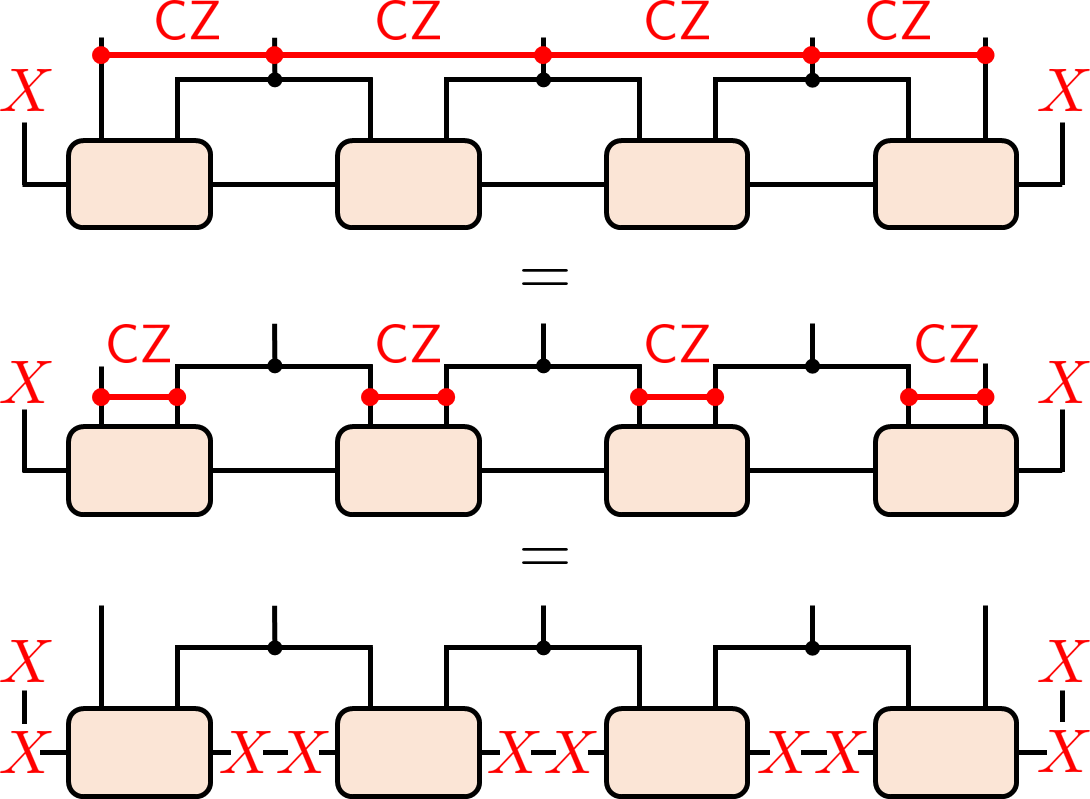}
    \caption{Demonstrating the global symmetry $\bar{U}_{\CZ}$~\eqref{eqn:UCZ-OBC} of the normal SIMPS with OBC. First, we use the trick in Eq.~\eqref{eqn:delta_trick} to pull the $\CZ$ operators down. Then, we use the second symmetry in Eq.~\eqref{eqn:nice-simps-symmetries} to push these onto the virtual legs, where the $X$'s cancel pairwise.}
    \label{fig:nice_simps_global_symm}
\end{figure}

Observe that the MPS matrices $A^0$ and $A^1$ in  Eq.~\eqref{eqn:nice-mps} both have rank two. Using Eq.~\eqref{eq:mpstosimps}, we can derive the equivalent SIMPS tensor with $\chi=2$,
\begin{align}\label{eqn:nice-SIMPS}
    B^{00} &= I  \quad \:\, B^{01} = I \notag \\
    B^{10} &= X \quad B^{11} = XZ
\end{align}%

On periodic boundaries, this SIMPS generates the same state as the MPS in Eq.~\eqref{eqn:nice-mps} up to normalization. Note that all $B^{ij}$ are unitary.
Later, we will view the tensor $B$ as a four-qubit quantum state, which we call $\ket{B}$. This state can be defined in terms of a quantum circuit,
\begin{equation}
    |B\rangle = \mathsf{CX}_{13}\mathsf{CCZ}_{123}|+\rangle_1|+\rangle_2|\Omega_2\rangle_{34}
    \, ,
    \label{eqn:SIMPS_circuit}
\end{equation}
where $\mathsf{CCZ} = I - 2|111\rangle\langle 111|$ is the controlled-controlled-$Z$ operator, and $\mathsf{CX}=|0\rangle\langle 0|\otimes I+ |1\rangle\langle 1|\otimes X$ is the controlled-$X$ operator. Graphically, this circuit is,
\begin{equation}
    \includegraphics[scale = 0.35]{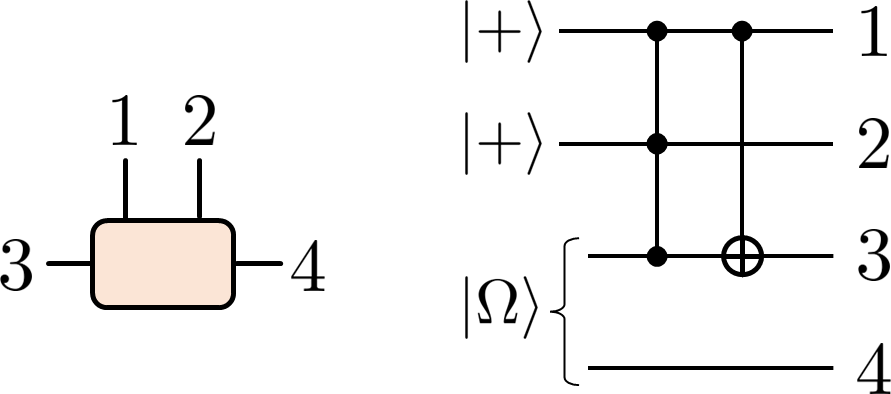}
\end{equation}

From this circuit representation, we can straightforwardly show that the tensor $B$ is invariant under the following four operations, which can be regarded as generalized push-through relations,
\begin{equation}
    \includegraphics[scale=0.17,raise=-0.05cm]{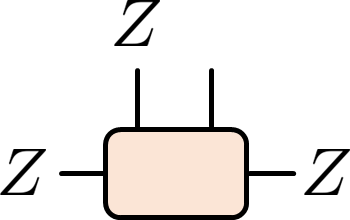}\ ,\hspace{1.7ex}
    \includegraphics[scale=0.17,raise=-0.05cm]{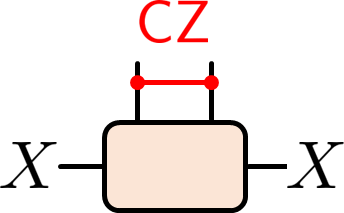}\ ,\hspace{1.7ex}
    \includegraphics[scale=0.17,raise=-0.05cm]{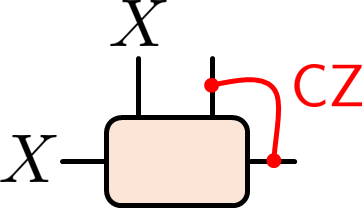}\ ,\hspace{1.7ex}
    \includegraphics[scale=0.17,raise=-0.05cm]{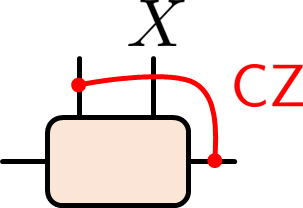}\ .
    \label{eqn:nice-simps-symmetries}
\end{equation}
These symmetries completely specify the state $|B\rangle$ up to a global phase. 
From the first two symmetries, we can derive a $\Z2\times\Z2$ global symmetry of the SIMPS. 
With open boundaries, the generating symmetries are 
\begin{align}
    \bar{U}_Z &= Z\otimes \prod_{k=1}^{N-1} Z_k\otimes Z, \\
    \bar{U}_{\CZ} &= X\otimes \prod_{k=1}^{N-1}\CZ_{k,k+1}\otimes X. \label{eqn:UCZ-OBC}
\end{align}
The symmetry $\bar{U}_{\CZ}$ is a non-onsite symmetry that acts on the bulk qudits as an FDQC. On PBC, $U_{\CZ}$ itself is a symmetry of the state. These symmetries follow from the symmetries of the local SIMPS tensor, as depicted in Fig.~\ref{fig:nice_simps_global_symm} for $\bar{U}_{\CZ}$. Indeed, one of the main strengths of SIMPS versus MPS is their ability to capture such non-onsite symmetries in the same way that MPS capture onsite symmetries \cite{SIMPS}. The fact that the actions of the two $\Z2$ symmetries on one of the boundary qudits anticommute indicates that this state has non-trivial SPT order under the $\Z2\times\Z2$ symmetry \cite{SIMPS}. Thus the situation is similar to the AKLT state, except now one of the symmetries is non-onsite. 

The non-normal example in Eq.~\eqref{eq:non_normal_mps} is similar. The corresponding SIMPS tensor is block diagonal in the virtual space,
\begin{align} \label{eqn:non_normal_SIMPS}
    B^{00} &= I  \quad B^{01} = I \notag \\
    B^{10} &= I \quad B^{11} = Z
\end{align}%
From these tensors it follows that Eq.~\eqref{eqn:SIMPS_circuit} is the same except the $\mathsf{CX}$ operator is absent and Eq.~\eqref{eqn:nice-simps-symmetries} is modified to,
\begin{equation}
    \includegraphics[scale=0.17,raise=-0.05cm]{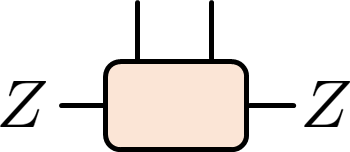}\ ,\hspace{1.7ex}
    \includegraphics[scale=0.17,raise=-0.05cm]{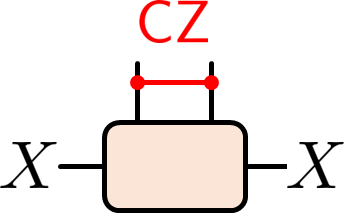}\ ,\hspace{1.7ex}
    \includegraphics[scale=0.17,raise=-0.05cm]{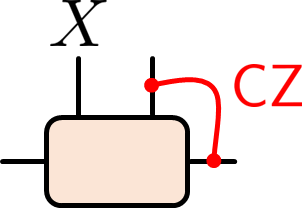}\ ,\hspace{1.7ex}
    \includegraphics[scale=0.17,raise=-0.05cm]{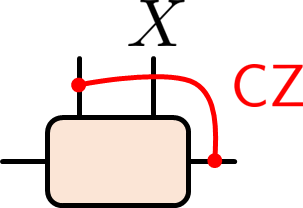}\ .
    \label{eqn:nonnormal-simps-symmetries}
\end{equation}
Notice that the leftmost symmetry has no $Z$ action on the upwards-pointing leg meaning virtual $Z$ operators can be pushed for free like in the GHZ state. Finally, $\bar{U}_{\CZ}$ is still a symmetry of the non-normal SIMPS, but the other $\Z2$ symmetry acts only as $Z$ on the two boundary qubits with no action in the bulk.


\section{Fusing SIMPS}
\label{sec:simps_fusion}

\begin{figure*}[t]
    \centering
    \includegraphics[scale=0.35]{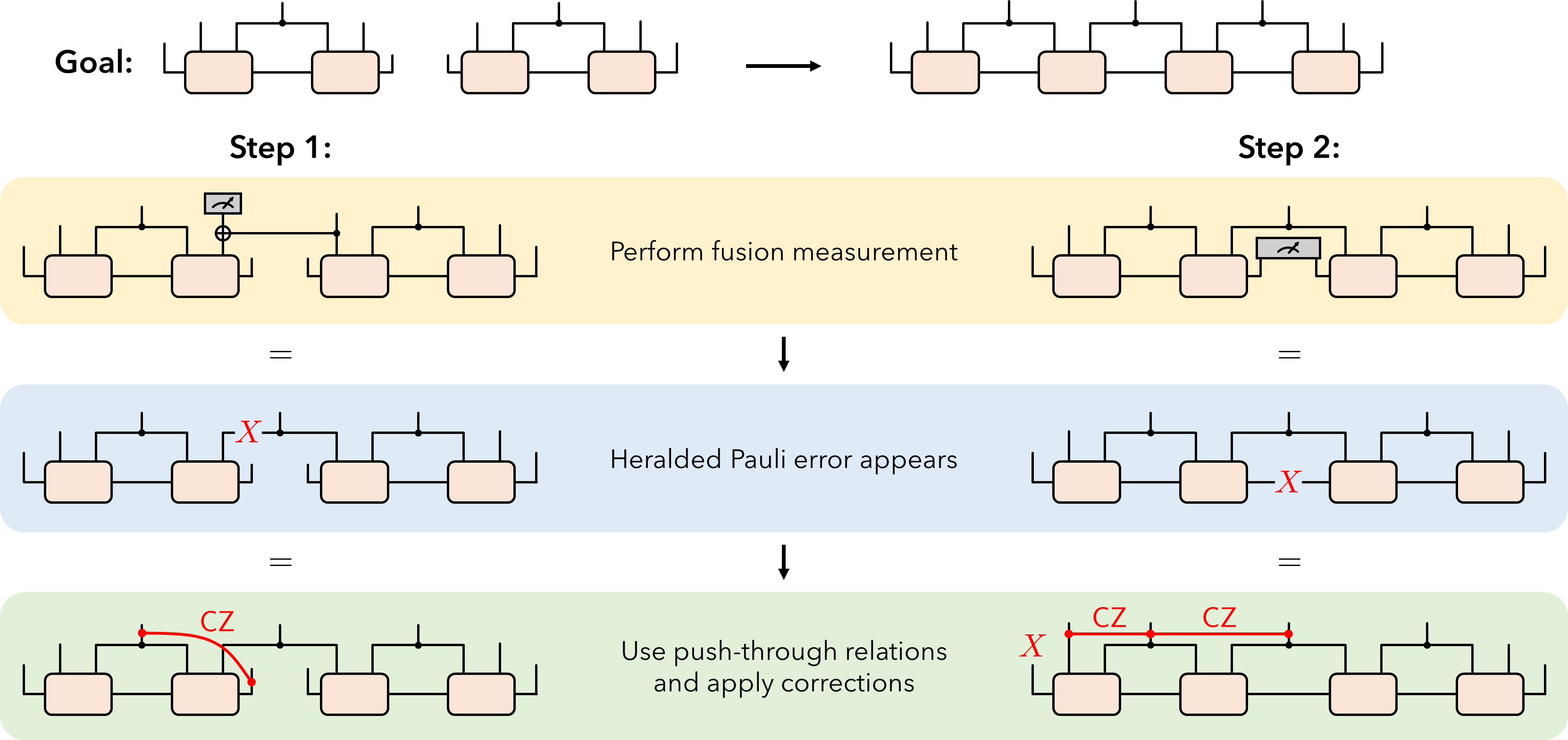}
    \caption{The protocol for fusing two copies of the normal SIMPS. We want to fuse two short SIMPS into a single longer one. The protocol has two steps. In the first step, we fuse the upper legs using a $\mathsf{CX}$ gate and a $Z$-basis measurement. When the measurement outcome is $s=1$, an $X$ operator is inserted into the tensor network as shown. The $X$ operator can be turned into a $\CZ$ using Eq.~\eqref{eqn:nice-simps-symmetries}, which can then be removed by using the trick in Eq.~\eqref{eqn:delta_trick} and acting with $\CZ$ on the appropriate legs. In the second step, we fuse the lower legs using a Bell measurement. Depending on the measurement outcome, a Pauli operator $P$ is inserted on the fused leg. For the case when $P=X$, this operator can be pushed to the boundary degree of freedom by applying the global symmetry $U_{\CZ}$ as in Fig.~\ref{fig:nice_simps_global_symm}. Undoing the $\CZ$ operators and the boundary $X$ completes the fusion. The case for $P=Z,Y$ are similar, except $P$ is pushed to the boundary using $Z$ operators or $Z$ and $\CZ$ operators, respectively. Note that the first step can also be implemented unitarily as discussed in the main text.}
    \label{fig:nice_simps_fusion}
\end{figure*}

Now, we show how the SIMPS representation leads to a deterministic protocol for creating the non-fusible MPS in constant depth with measurements and feedforward. The key idea is as follows. Since the SIMPS representation effectively has two legs connecting each tensor [see Eq.~\eqref{eq:obc_simps}], the fusion protocol consists of two steps where we first merge the upper legs connecting each tensor and then the lower legs. The protocol for the normal SIMPS is shown in Fig.~\ref{fig:nice_simps_fusion}. The protocol for the non-normal SIMPS is identical, except that byproduct $Z$ operators are pushed to the boundary for free, rather than by using bulk $Z$'s as in the normal case. We only discuss preparing the SIMPS on OBC. As for MPS, we can prepare the SIMPS on PBC by fusing the legs at the two boundaries of the SIMPS on OBC and postselecting on the outcome that has no Pauli byproduct.

We note two distinct features of SIMPS fusion as compared to MPS fusion. First, the protocol involves two distinct rounds of measurement where the correction unitaries resulting from the first round of measurements must be applied before the second round of measurements are performed. These measurements also occur on $\chi=2$ dimensional qudits rather than the $D=3$ dimensional qudits that would be measured in MPS fusion. Second, the correction unitaries are not always single-site operators and instead involve two-site entangling unitaries as well. This is a consequence of the non-onsite symmetries of the SIMPS, which we are employing to push byproduct operators through the SIMPS virtual space (Fig.~\ref{fig:nice_simps_global_symm}). 
Practically, the application of entangling correction unitaries does not drastically affect the resource requirements of the procedure since entangling operations between degrees of freedom are already required to perform the Bell measurements.

The SIMPS fusion in these examples is non-generic in that the classical communication required in the first step is local, meaning each correction unitary only depends on nearby measurement outcomes. This should be compared to the non-local classical communication that is necessary in MPS fusion and the second step of SIMPS fusion, i.e., the correction unitaries can depend on the outcomes of arbitrarily distant measurements. In general, the first step of SIMPS fusion may also require non-local classical communication. This occurs, for example, when SIMPS fusion is used to construct GHZ states, as discussed in Sec.~\ref{sec:comparison}. It should be noted that some degree of non-local classical communication is necessary if one wants to prepare states with non-zero correlation length in constant depth~\cite{friedman2023locality}.

Finally, we remark that, because the first step of SIMPS fusion in Fig.~\ref{fig:nice_simps_fusion} requires only local classical communication, it can in fact be done unitarily, i.e., without measurements. Rather than measuring the qubit in the first step, we can apply a controlled unitary that implements the necessary correction unitary depending on the state of that qubit. In the particular case described in Fig.~\ref{fig:nice_simps_fusion}, this amounts to applying a $\mathsf{CCZ}$ operator to the qubit that would be measured and the two qubits on which the $\mathsf{CZ}$ correction unitary acts. Since the protocol is deterministic, this will result in the desired fusion of SIMPS legs, where the qubit that would have been measured ends up in a disentangled product state. In the more general case where non-local classical communication is necessary, these controlled unitaries would be long-range and potentially many-body, making them infeasible to implement, so using measurements is preferable.


\subsection{Fusing anomalous SIMPS}
\label{sec:omega_simps}

Here we describe a generic construction of SIMPS that are invariant under anomalous symmetries, and show how to prepare them using SIMPS fusion. The purpose of this construction is to expand the scope of SIMPS fusion, and to highlight its relation to the physics of lattice anomalies.

First, we define a general class of non-onsite symmetries that form a representation of a group $G$. To any such symmetry, one can associate an invariant called a 3-cocycle $\omega$ which is a function $\omega:G\times G\times G\to U(1)$ that satisfies a certain ``cocycle condition'' \cite{Chen2011,Else2014}, see Appendix \ref{app:anom_calc} for details. The 3-cocycles can be organized into equivalence classes called cohomology classes, and the different classes correspond to different lattice anomalies in 1D. In the present context, an anomaly manifests as an obstruction to finding a short-range-entangled state that is invariant under the symmetry \cite{Chen2011}.

In Ref.~\cite{GarreRubio2023classifyingphases}, the authors constructed examples of non-onsite symmetries for any group $G$ and cocycle $\omega$. The symmetries are defined on a Hilbert space spanned by the states $\{|g\rangle\}_{g\in G}$ as follows,
\begin{equation}
    U^\omega(g) = \prod_i L(g)_i \prod_i u^\omega(g)_{i,i+1}\ .
\end{equation}
where $L(g)=\sum_{h\in G} |gh\rangle\langle h|$ permutes the states $\{|h\rangle\}_{h\in G}$ by performing left multiplication by $g$ and $u^\omega(g)= \sum_{h,k\in G} \omega(g,h,h^{-1}k)|h,k\rangle\langle h,k|$.
For simplicity, we restrict to finite abelian $G$. We also restrict to cocycles that are linear in the first input, meaning that $\omega(g,h,k)\omega(g',h,k)=\omega(gg',h,k)$. We note that every cohomology class contains (at least) one cocycle satisfying this condition (see, \textit{e.g.}, those defined in Ref.~\cite{propitius1995topologicalinteractionsbrokengauge}), so this linearity assumption does not limit the kinds of anomalies we can capture.  Using linearity, it is clear that $g\mapsto u^\omega(g)$ forms a representation of $G$. As we show in Appendix \ref{app:anom_calc}, $\prod_i L(g)_i$ and $\prod_i u^\omega(g)_{i,i+1}$ commute with each other and therefore form a representation of $G\times G$.

The authors of Ref.~\cite{GarreRubio2023classifyingphases} also classified phases of matter protected by such symmetries, and gave representative fixed-point states for each. Here, we focus on a particular class of symmetric states, but we expect that all fixed-point states defined in Ref.~\cite{GarreRubio2023classifyingphases} will be amenable to SIMPS fusion or a straightforward generalization thereof.
We consider the following family of (unnormalized) states,
\begin{equation}
    |\psi_\omega\rangle = \sum_{g\in G} \prod_i u^\omega(g)_{i,i+1} \ket{+_G+_G\dots +_G}\ ,
\end{equation}
where we defined $\ket{+_G} = \sum_{g\in G} |g\rangle /\sqrt{|G|}$.
Thanks to the properties stated above, the state $|\psi_\omega\rangle$ is invariant under $\prod_i L(g)_i$ and $\prod_i u^\omega(g)_{i,i+1}$ independently, and hence also invariant under $U^\omega(g)$ for all $g\in G$.

The state $|\psi_\omega\rangle$ can be represented using the following SIMPS tensor,
\begin{equation} \label{eq:omega_simps}
    \includegraphics[scale=0.35,valign=c]{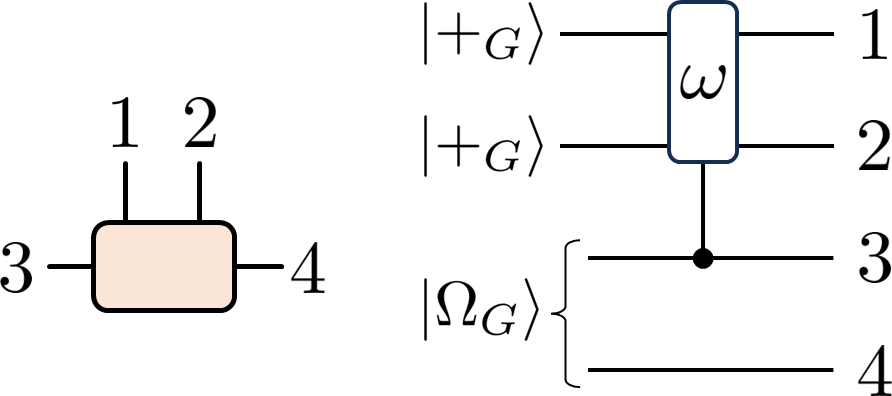}
\end{equation}
where $|\Omega_G\rangle = \sum_{g\in G} |gg\rangle /\sqrt{|G|}$ and the controlled unitary labeled by $\omega$ denotes the operator $\sum_g u^\omega(g)\otimes |g\rangle\langle g|$. As the simplest non-trivial example, taking $G=\mathbb{Z}_2=\{e,x\}$ and choosing $\omega$ such that $\omega(x,x,x)=-1$ and all other entries are equal to one, we have $U^\omega(e)=I$ and $U^\omega(x)=\prod_i X_i \prod_i Z_{i}\CZ_{i,i+1}$, such that,
\begin{equation}
    |\psi_\omega\rangle = |++\cdots +\rangle + \prod_i Z_i \CZ_{i,i+1}|++\cdots +\rangle\ .
\end{equation}
This state is very similar to the non-normal SIMPS defined earlier in Eq.~\eqref{eq:non_normal_mps}, and it shares the property of having a non-degenerate entanglement spectrum. This state also appears as an example of the classification of phases with anomalous symmetries in Ref.~\cite{GarreRubio2023classifyingphases}.

Because of the above example, we see that not all of the states $|\psi_\omega\rangle$ are preparable using MPS fusion. A simple argument shows that, if $|\psi_\omega\rangle$ is preparable using MPS fusion, then $\omega$ must belong to a trivial cohomology class, \textit{i.e.}, there is no anomaly. Indeed, in Sec.~\ref{sec:mps_fusion_constraints} we showed that MPS fusibility implies that the state is a sum of normal MPS, each of which has the same bond dimension. Now, $|\psi_\omega\rangle$ is a sum of normal MPS $\prod_i u^\omega(g)_{i,i+1}|+_G+_G\dots +_G\rangle$, including the case $g=e$, which is a product state since $u^\omega(e)=I$. Therefore, the only way for each of these MPS to have the same bond dimension is if they are all product states, \textit{i.e.}, if $u^\omega(g)$ is a product of single-site unitaries for all $g$. However, onsite representations of $G$ always belong to the trivial cohomology class \cite{Else2014}, so there is no anomaly.

Now we show how to prepare each of the states $|\psi_\omega\rangle$ using SIMPS fusion, including those for which $\omega$ is anomalous. The fusion procedure for $|\psi_\omega\rangle$ is the same as in Fig.~\ref{fig:nice_simps_fusion}, with the various measurements and correction operators replaced with generalized operators defined by $G$ and $\omega$. More precisely, using the linearity of $\omega$ in its first input, we can derive the following symmetries of the SIMPS tensor \eqref{eq:omega_simps},
\begin{equation}
    \includegraphics[scale=0.34,raise=-0.05cm]{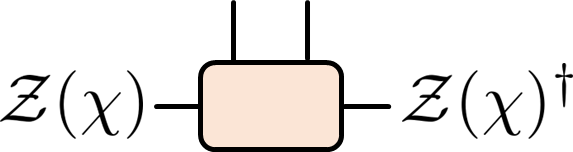}\ ,\hspace{1.7ex}
    \includegraphics[scale=0.34,raise=-0.05cm]{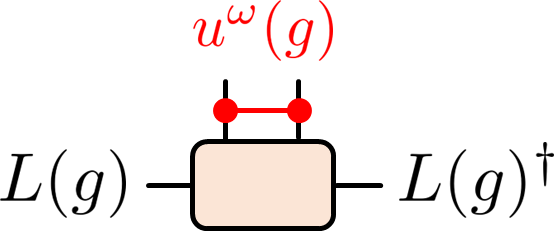}\ ,\hspace{1.7ex}
    \includegraphics[scale=0.34,raise=-0.05cm]{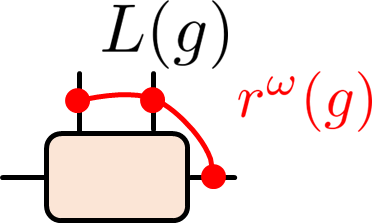}\
    \label{eqn:omega-simps-symmetries}
\end{equation}
where $\mathcal{Z}(\chi) = \sum_{g\in G} \chi(g)|g\rangle\langle g|$ for $\chi:G\rightarrow U(1)$ a linear character of $G$. Note that, since $G$ is finite abelian, the linear characters form an abelian group $\hat{G}$ that is isomorphic to $G$. The three-body unitary $r^\omega(g)$ is a diagonal unitary defined by $\omega$ whose precise form is given in Appendix \ref{app:anom_calc}. To generalize Step 1 of the fusion protocol shown in Fig.~\ref{fig:nice_simps_fusion}, the $\CX$ gate is replaced by the controlled-left multiplication operator $\sum_g |g\rangle\langle g|\otimes L(g)^\dagger$ and the measurement is performed in the group basis $\{|g\rangle\}$. The resulting byproduct is generalized from $X$ to $L(g)$ depending on the measurement outcome labeled by $g$, and the byproduct is removed by using the last symmetry in Eq.~\eqref{eqn:omega-simps-symmetries} and applying $r^\omega(g)$ instead of $\CZ$. Note that, since $r^\omega(g)$ is a generically a three-body unitary, it needs to be applied to three, rather than two, open legs of the SIMPS (the third leg being the central leg in bottom left diagram of Fig.~\ref{fig:nice_simps_fusion}).
To generalize Step 2, the fusion measurement is performed in the basis $\{ (\mathcal{Z}(\chi) L(g)\otimes I)|\Omega_G\rangle \,|\, g\in G,\chi\in\hat{G}\}$ labeled by group elements $g\in G$ and characters $\chi\in \hat{G}$. There are $|G|^2$ such states, all of which are orthogonal thanks to the orthogonality of linear characters, so these states define a valid basis for the fusion measurement. Then, the byproduct operator after the fusion measurement is $\mathcal{Z}(\chi) L(g)$, which can be pushed to the boundary leg (and subsequently annihilated) by applying $u^\omega(g)$ according to the symmetries in Eq.~\eqref{eqn:omega-simps-symmetries}.

\subsection{Constraints on SIMPS fusion} \label{sec:simps_constraints}

Here, we outline some constraints on SIMPS fusion, similar to those that we have derived for MPS fusion. In particular, under some assumptions on the form of the correction unitaries, which hold for all examples we have considered, we prove that states that can be created using SIMPS fusion have a sort of hidden entanglement-spectrum degeneracy. This hidden degeneracy was previously observed for the case of the normal SIMPS \eqref{eqn:nice-SIMPS} in Ref.~\cite{SIMPS}, and here we prove that this is a necessary property of SIMPS-fusible states.

The constraint comes from the second step of SIMPS fusion. That is, we assume that the first step in Fig.~\ref{fig:nice_simps_fusion} has been performed, and we determine necessary conditions for the second step to be possible. Suppose that the following transformation can be performed using a fusion measurement,
\begin{equation}
    \includegraphics[scale=0.34,valign=c]{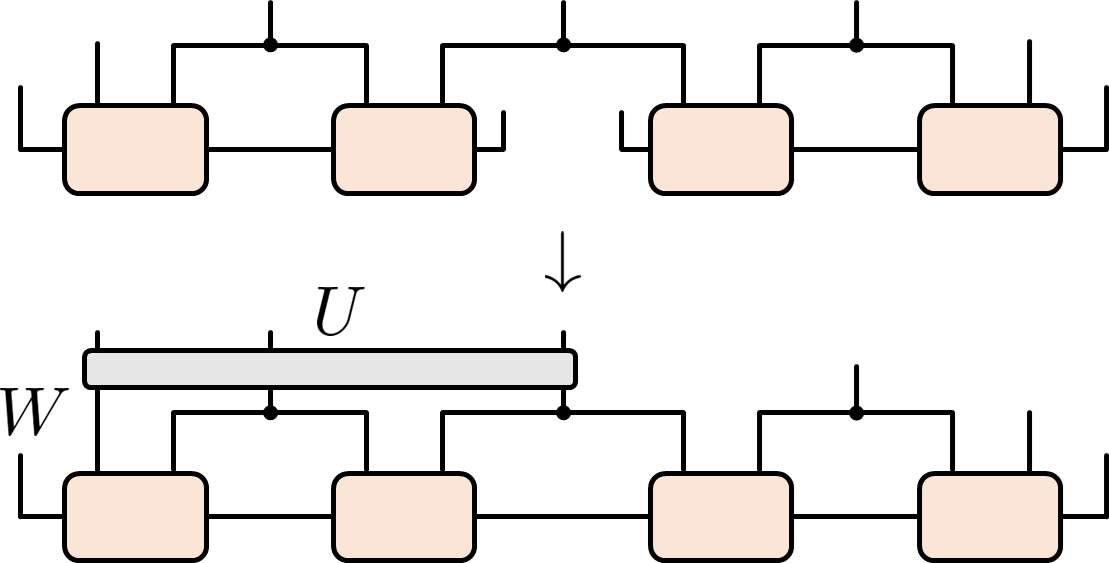}
    \label{eqn:simps_fusion_step2}
\end{equation}
where $U$ and $W$ are unitary correction operators, generalizing Step 2 in Fig.~\ref{fig:nice_simps_fusion}. Suppose further that $U$ is diagonal in the computational basis (the basis in which the $\delta$-tensor \eqref{eqn:delta-tensor} is defined), as was the case for all SIMPS fusion protocols described up to this point. Now, we take Eq.~\eqref{eqn:simps_fusion_step2} and project every other $d$-dimensional qudit onto a fixed-state $|s\rangle$ in the computational basis. This gives,
\begin{equation}
    \includegraphics[scale=0.34,valign=c]{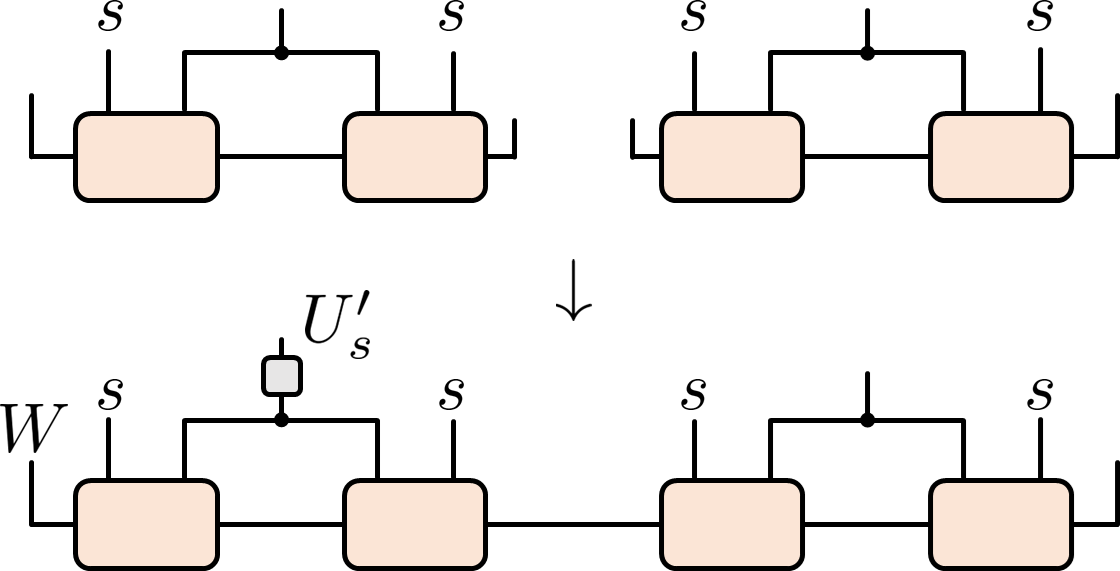}
    \label{eqn:simps_fusion_step2_proj}
\end{equation}
where $U'_s$ is defined by projecting the unitary $U$ onto the states $|s\rangle$ on the appropriate legs, and it is unitary because we assumed $U$ is diagonal \footnote{In a more general situation, $U'_s$ may also act in a non-onsite manner on the unmeasured legs, but this case is covered by the more general assumptions in Appendix \ref{app:asymmetric-pushthrough}}. Now, Eq.~\eqref{eqn:simps_fusion_step2_proj} describes the fusion of an MPS defined by the tensor $A_s^i=B^{si}B^{is}$ where $B^{ij}$ is the original SIMPS tensor. By the results of Sec.~\ref{sec:mps_fusion_constraints}, the state described by $A^i_s$ must have a flat entanglement spectrum for all $s$. Therefore, the ability to prepare a state using SIMPS fusion (subject to the assumptions above) implies that projecting every other qudit onto a fixed state results in a state that has a degenerate entanglement spectrum. One can verify that this hidden degeneracy is indeed present in all examples considered until now. For example, projecting every other qubit onto $|0\rangle$ in the normal SIMPS \eqref{eqn:nice-mps} leaves the state $\ket{++\cdots +}$ on the remaining spins, while projecting onto $|1\rangle$ leaves a GHZ state. We note that this is not the only way of revealing the hidden entanglement degeneracy. For example, it is shown in Ref.~\cite{SIMPS} that it can be revealed by fixing the state of a single qudit next to the entanglement cut.

Importantly, we have not commented at all on whether the first step of SIMPS fusion is possible, so we suspect that this hidden degeneracy, which comes only from the second step, is far from a sufficient condition for SIMPS fusibility.


\section{General framework for fusion}

Here, we sketch a more general framework for the fusion of 1D quantum states and explain how MPS fusion and SIMPS fusion fit into this framework. 
Then, we compare the resource requirements for MPS and SIMPS fusion in cases where both protocols are capable of preparing a particular quantum state.

\subsection{Unifying MPS and SIMPS fusion}
\label{sec:unifying_framework}

A general fusion protocol should use a finite-depth circuit consisting of local measurements and unitary feedforward, perhaps in several rounds, to take two copies of a state and produce a single copy that is approximately twice as large,
\begin{equation} \label{eqn:general_fusion}
    |\psi_L\rangle \otimes |\psi_L\rangle \rightarrow |\psi_{2L+C}\rangle
\end{equation}
where the integer $C \geq 0$ does not depend on $L$ (\textit{e.g.}~Figs.~\ref{fig:aklt_fusion} and \ref{fig:nice_simps_fusion} show that MPS fusion and SIMPS fusion have $C=0$ and $C=1$, respectively). The depth of the circuit and the number of rounds of measurement should also be independent of $L$. More generally, we could allow ancillas on either side of Eq.~\eqref{eqn:general_fusion}, but the number of ancillary qubits should scale at most linearly with $L$.

\begin{figure}[t]
    \centering
    \includegraphics[scale=0.35]{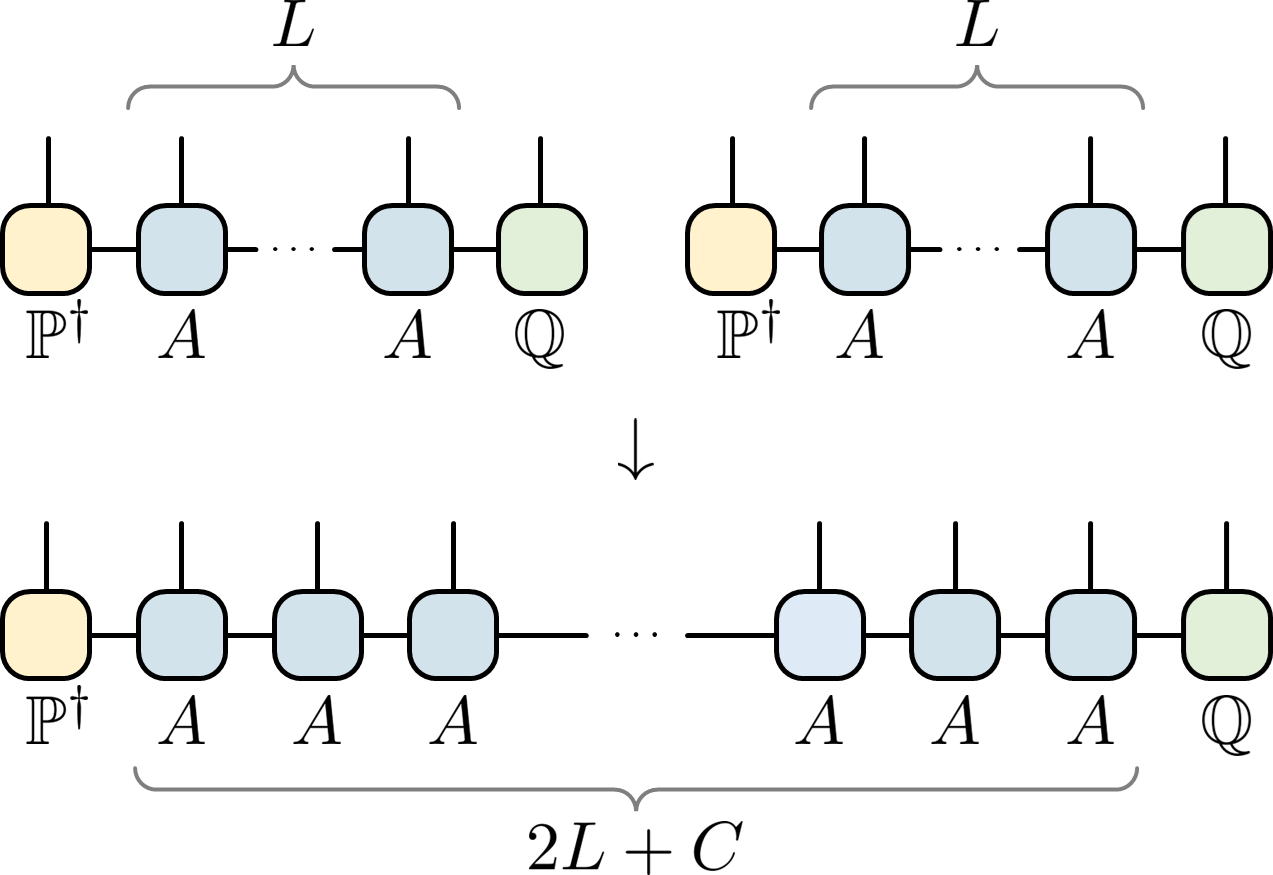}
    \caption{Generalized fusion of MPS which encompasses both MPS fusion and SIMPS fusion. Two states of size $L$ with boundary conditions defined by $\mathbb{P}$ and $\mathbb{Q}$ are fused to a state of size $2L+C$ with the same boundary conditions.}
    \label{fig:gen_fusion}
\end{figure}

Importantly, this definition requires a choice of how to define the state $|\psi_L\rangle$ with open boundary conditions. The MPS representation gives a natural way of defining a state for any length $L$, and therefore is a natural tool for studying fusion in general. We define the state $|\psi_L\rangle$ with general boundary conditions as,
\begin{equation} \label{eq:obc_mps_gen}
\begin{aligned}
     |\psi_L\rangle = \sum_{a,b=0}^{D'-1} \sum_{i_1,\dots,i_L=0}^{d-1} &\langle a| \mathbb{P}^\dagger A^{i_1}A^{i_2}\cdots A^{i_L} \mathbb{Q}|b\rangle \\
     &\times |a,i_1 i_2\cdots i_L,b\rangle
\end{aligned}
\end{equation}
where the state is terminated with $D'$-dimensional qudits at each boundary and the matrices $\mathbb{P}$ and $\mathbb{Q}$ are maps from $\mathbb{C}^{D'}$ to $\mathbb{C}^D$ that determine the boundary conditions, where $D$ is the bond dimension. For $|\psi_L\rangle$ defined in this way, the fusion process described by Eq.~\eqref{eqn:general_fusion} is pictured in Fig.~\ref{fig:gen_fusion}. We stress that, even though it is convenient to express $|\psi_L\rangle$ as an MPS, this does not mean that any fusion protocol is equivalent to MPS fusion, as originally defined in Ref.~\cite{Smith_AKLT}. Indeed, we will soon see that SIMPS fusion falls under this framework, but SIMPS fusion goes beyond MPS fusion as we have demonstrated.

The OBC MPS \eqref{eqn:MPS} has the form of Eq.~\eqref{eq:obc_mps_gen} with $D'=D$ and $\mathbb{P}=\mathbb{Q}=I$. 
We can also put the OBC SIMPS \eqref{eq:obc_simps_eq} into this form as follows. Using Eq.~\eqref{eq:mpstosimps}, we can rewrite Eq.~\eqref{eq:obc_simps_eq} as,
\begin{equation}
    \ket{\bar{\psi}} = \sum_{\substack{i_1, \dots, i_N \\ a, b}} \braket{a | P^{i_1 \dagger}A^{i_2} A^{i_3} \cdots A^{i_N}P^{i_N} | b} \ket{a,i_1 i_2 \cdots i_N,b}
    \, ,
\end{equation}
By considering qudits $i_1$ and $i_N$ as part of the boundary degrees of freedom, this has the form of Eq.~\eqref{eq:obc_mps_gen} with $L=N-2$, $D'=d\chi$, $\mathbb{P} = \sum_{i=0}^{d-1} \langle i|\otimes P^i$ and $\mathbb{Q}= \sum_{i=0}^{d-1} \langle i|\otimes A^i P^i$. Therefore, an OBC SIMPS can alternatively be viewed as an OBC MPS with a particular choice of boundary condition. Notably, this boundary condition is very unnatural from the MPS point of view, whereas it is the most natural boundary condition when using the SIMPS representation.

From the perspective of this general framework, SIMPS fusions highlights how the basic MPS fusion protocol \cite{Smith_AKLT} can be extended to broaden the classes of states that can be prepared via fusion. Namely, SIMPS fusion shows the importance of 
\begin{enumerate*}[label=(\roman*)]
    \item multiple rounds of measurement and feedforward,
    \item correction operations beyond single-site unitaries, and
    \item different choices of boundary condition.
\end{enumerate*}
Moving forward, we hope that this framework can be used to further expand the class of fusible states and potentially lead to a full classification. In the latter direction, we note that both SIMPS and MPS fusion are associated with a (hidden) entanglement-spectrum degeneracy, so it is interesting to ask whether this is always the case.

\subsection{Comparing MPS and SIMPS fusion} \label{sec:comparison}

Having seen that they are both instances of a larger framework, we now make comparisons between MPS and SIMPS fusion. First, we remark that SIMPS fusion is strictly more powerful than MPS fusion. This is because any MPS defined by the tensor $A^i$ can be written as a SIMPS with $B^{ij}=A^i$, although this will in general not give the minimal SIMPS bond dimension, which would be provided by Eq.~\eqref{eq:mpstosimps}. If MPS fusion is possible for the tensor $A^i$, then it is straightforward to see that SIMPS fusion will also be possible for $B^{ij}=A^i$, although the first step of SIMPS fusion is essentially pointless in this case, resulting in wasted resources. Nevertheless, it stands that SIMPS fusion is possible whenever MPS fusion is. 

In certain cases, SIMPS fusion and MPS fusion are both possible, but SIMPS fusion requires arbitrarily fewer ancilla qubits.
In general, MPS fusion requires measuring two $D$-dimensional ancilla qudits per site, whereas SIMPS fusion requires measuring two $\chi$-dimensional qudits but only one $d$-dimensional qudit per site (since one of the $d$-dimensional qudits becomes part of the final state). Therefore, if we measure complexity of the fusion by the size of the ancillary Hilbert space needed for each fusion, SIMPS fusion outperforms MPS fusion whenever $d\chi^2 < D^2$.

We define the quantity $\Theta = \log_2 D^2-\log_2 d\chi^2$, which counts the difference in the number of ancilla qubits per site needed for MPS fusion versus SIMPS fusion (assuming that both are possible).
For almost all SIMPS with physical dimension $d$ and bond dimension $\chi$, the corresponding MPS bond dimension will be $D=d\chi$. Therefore, we generically have $D^2=d^2\chi^2$ such that $\Theta = \log_2 d$. Since $d$ can be arbitrarily large, SIMPS fusion can be arbitrarily better than MPS fusion in terms of the number of ancillary qubits. Conversely, for a generic MPS of bond dimension $D$, the SIMPS bond dimension will be $\chi=D$, such that $\Theta = -\log_2 d$ and MPS fusion is more resource efficient. This suggests the intuitive idea that generic MPS are more suited to MPS fusion whereas generic SIMPS are more suited to SIMPS fusion. 

The above argument is somewhat tenuous since generic MPS and SIMPS are not fusible, and it may be the case that the relationship between the bond dimensions of generic MPS and SIMPS is different in the case that both are fusible. Nevertheless, we can realize the large gap between MPS and SIMPS fusion explicitly with qudit GHZ states defined as,
\begin{equation}
    |\text{GHZ}_d\rangle = \frac{1}{\sqrt{d}}\sum_{i=0}^{d-1}|ii\cdots i\rangle.
\end{equation}
These can be realized as MPS with bond dimension $D=d$ defined by the matrices $A^i=|i\rangle\langle i|$. Then, MPS fusion uses two ancillary qudits per site. Instead, we can write $|\text{GHZ}_d\rangle$ as a $\chi=1$ SIMPS defined by the tensor $B^{ij}=\delta_{ij}$. Note that $D=d\chi$ (the generic case) holds here. Since $\chi=1$, the second step of SIMPS fusion is non-existent. Therefore, SIMPS fusion uses only one ancillary qudit per site, so $\Theta = \log_2 d$. For $d=2$, this reproduces the resource overhead of the preparation scheme of Ref.~\cite{tantivasadakarn2022longrange}. We remark that, in this case, unlike the examples in Sec.~\ref{sec:simps_fusion}, the first step of SIMPS fusion for the GHZ state involves non-local classical communication. Another family of states with $\Theta=\log_2 d$ is given by qudit cluster states \cite{Zhou2003}, although these are somewhat trivial since they can be constructed in constant depth without measurements.


\section{Discussion}

We presented constraints on which states can be deterministically prepared using fusion measurements on MPS. The main constraint can be summarized by the necessity of a flat entanglement spectrum. We then showed how to go beyond this constraint by instead expressing states in the SIMPS representation. This leads to a new preparation protocol which has novel properties compared to MPS fusion such as the necessity of multiple rounds of measurement and unitary corrections that have the form of FDQCs. The examples we prepared with SIMPS fusion are associated with symmetry-breaking and SPT order protected by non-onsite symmetries, and we argued that MPS fusion cannot prepare states with these physical properties, meaning that our protocol unlocks the preparation of new physical phases of matter. {We also proposed a framework for fusion that includes both MPS fusion and SIMPS fusion. We hope this framework will lead to a classification of fusible 1D states in the most general case.}

One natural extension of our results is to consider preparing higher-dimensional tensor networks like projected entangled pair states (PEPS) and the multiscale entanglement renormalization ansatz (MERA) with fusion. This has already been considered for some simple topological PEPS in Refs.~\cite{Lu2022Measurement,Herringer2023classificationof}. When moving to more general models like string-net PEPS, the push-through relations are expressed in terms of matrix product operators \cite{Sahinoglu2021,Cirac2021}, and it is not clear how these relations can be leveraged to perform deterministic fusion of legs in the PEPS. We remark that it is still an open question whether string nets like the doubled Fibonnaci model can be deterministically created in constant depth with measurements \cite{Tantivasadakarn2023hierarchy}, and the generalized perspective on fusion that we have presented herein may be helpful there.

Finally, while we focused on deterministic state preparation, it is also interesting to consider the properties of the ensembles of states that result from different measurement outcomes. These so-called projected ensembles are useful for generating quantum state designs \cite{Choi2023,Cotler2023,Ippoliti2023} and realizing exotic kinds of entanglement and criticality \cite{Zhu2023,lee2022decoding,Lavasani2023,Lu2023mixed,Garratt2023}. It would be interesting to formulate some of these phenomena in terms of ensembles of tensor network states resulting from fusion measurements.

\emph{Note added.}---%
The posting of this preprint to the arXiv was coordinated with simultaneous postings by Sahay and Verresen~\cite{RahulGeneralFusion1,RahulGeneralFusion2} and Smith \emph{et al.}~\cite{SmithGeneralFusion}. Both discuss the measurement-based preparation of matrix product states, and were developed independently from the work here. {Our results agree where they overlap. In particular, Ref.~\cite{RahulGeneralFusion2} also derives entanglement-spectrum constraints on MPS fusion, while Ref.~\cite{SmithGeneralFusion} also goes beyond these constraints by allowing multiple rounds of measurement.}

\begin{acknowledgments}
OH is supported by the Air Force Office of Scientific Research under Grant FA9550-20-1-0222. DTS is supported by the Simons Collaboration on Ultra-Quantum Matter, which is a grant from the Simons Foundation (651440).
\end{acknowledgments}

\appendix

\section{Entanglement spectrum of OBC MPS} \label{sec:ent_spec}

Here we derive a formula for the entanglement spectrum of the OBC MPS defined in Eq.~\eqref{eqn:MPS}, following Ref.~\cite{Cirac2011}. 
Recall that the most general MPS that we attempt to fuse are \emph{block injective}:
\begin{equation}
    A^i = \bigoplus_{\alpha=1}^{N_b} A_\alpha^i
    \label{eqn:Ai-block-injective}
\end{equation}
where $\Span_{i_1,\dots,i_L} A^{i_1}\cdots A^{i_L}$ is equal to the space of all block-diagonal matrices with the same block structure as Eq.~\eqref{eqn:Ai-block-injective} for $L$ larger than some $L_0$ called the injectivity length~\cite{Cirac2021}. 
The OBC MPS can be written as,
\begin{equation}
    \ket{{\psi}} = 
    \sum_{\substack{i_1, \dots, i_N \\ a,b}} 
     \braket{a | \bigoplus_\alpha A_\alpha^{i_1} A_\alpha^{i_2} \cdots A_\alpha^{i_{N}} | b} \ket{a, i_1, i_2, \dots, i_N, b}
    \, .
\end{equation}
If we label the basis states of the boundary Hilbert space by the blocks $\alpha$, such that $a$ takes values $\alpha_{1},\dots,\alpha_{D_\alpha}$ for all $\alpha$, then we can write,
\begin{equation}
    |{\psi}\rangle = \bigoplus_{\alpha=1}^{N_b}  |{\psi}_\alpha\rangle
\end{equation}
where, for each block, we have,
\begin{equation}
    \ket{{\psi}_\alpha} = 
    \sum_{\substack{i_1, \dots, i_N \\ a,b}} 
    \braket{a | A_\alpha^{i_1} A_\alpha^{i_2} \cdots A_\alpha^{i_{N}} | b} \ket{a, i_1, i_2, \dots, i_N, b}
    ,
\end{equation}
where $a,b$ now run over the restricted set $\alpha_1,\dots,\alpha_{D_\alpha}$. The direct sum structure appears here due to the boundary particles; each state $|\psi_\alpha\rangle$ is supported in a different sector of the boundary Hilbert space.
Note that each $\ket{{\psi}_\alpha}$ is a normal MPS.
Now, we can write,
\begin{equation} \label{eq:rho-A-decomp}
    \rho^A=\bigoplus_{\alpha=1}^{N_b} \rho^A_\alpha
    \, ,
\end{equation}
where $\rho^A_\alpha = \Tr_B |\psi_\alpha\rangle\langle \psi_\alpha|$ for some spatial bipartition of the system into contiguous subsystems $A$ and $B$. Since each state $|\psi_\alpha\rangle$ is a normal MPS, one can show that,
\begin{equation}
    \spec (\rho_\alpha^A) \simeq \spec \left( \sqrt{\sigma_{L_\alpha}^T}\sigma_{R_\alpha}\sqrt{\sigma_{L_\alpha}^T} \right)
\end{equation}
where $\simeq$ denotes that the two sides have identical non-zero eigenvalues.
The matrices $\sigma_{L_\alpha}$, $\sigma_{R_\alpha}$ are the left and right fixed points, respectively, of the transfer matrix corresponding to the MPS tensor $A^i_\alpha$ \cite{Cirac2011}. Since $\sigma_{L_\alpha}$ is invertible, $\rho^A_{\alpha}$ has the same non-zero eigenvalues as $\sigma_{R_\alpha}\sigma_{L_\alpha}^T$. Therefore, the full reduced density matrix $\rho^A$ shares its non-zero eigenvalues with
\begin{equation} \label{eq:sigma-decomp}
    \sigma = \bigoplus_{\alpha=1}^{N_b} \sigma_{R_\alpha}\sigma_{L_\alpha}^T\, .
\end{equation}
Thus, the non-zero eigenvalues of the entanglement spectrum are encoded in $\sigma$.

Equation~\eqref{eq:sigma-decomp} is exactly $\tilde{\Pi}$ defined below Eq.~\eqref{eq:partial_projector}.
To see this, we write the MPS transfer matrix as
\begin{equation}
     T = \bigoplus_{\alpha,\beta} T_{\alpha\beta}
\end{equation}
where, in terms of the blocks in Eq.~\eqref{eq:A-decomp},
\begin{equation}
    T_{\alpha\beta} = \sum_i A^i_\alpha\otimes \bar{A}^i_\beta\, .
\end{equation}
Note that, for a PBC system of size $N$, we have $\langle \psi_\alpha|\psi_\beta\rangle = \Tr T^N_{\alpha\beta}$. Using the assumed linear independence of the states generated by different blocks ($\alpha \neq \beta$), we obtain that this overlap is $\delta_{\alpha\beta}$ for $N\to\infty$, and it follows that the matrices $T_{\alpha\beta}$ must have leading eigenvalue strictly less than 1 for $\alpha\neq\beta$. Therefore, we can write the fixed-point projector as,
\begin{equation} \label{eq:projector_decomp}
    \Pi = \bigoplus_{\alpha,\beta} \delta_{\alpha\beta} \Pi_\alpha ,
\end{equation}
where $\Pi_\alpha=|R_\alpha\rangle\langle L_\alpha|$ is the projector onto the unique fixed-points of the block $T_{\alpha\alpha}$. This gives,
\begin{equation} \label{eq:partial_trace_Pi}
    \tilde{\Pi} = \bigoplus_\alpha \sigma_{R_\alpha}\sigma_{L_\alpha}^T = \sigma ,
\end{equation}
as claimed.

\section{Generalized push-through relations}
\label{app:asymmetric-pushthrough}

\begin{figure}
    \centering
    \includegraphics[scale=0.175]{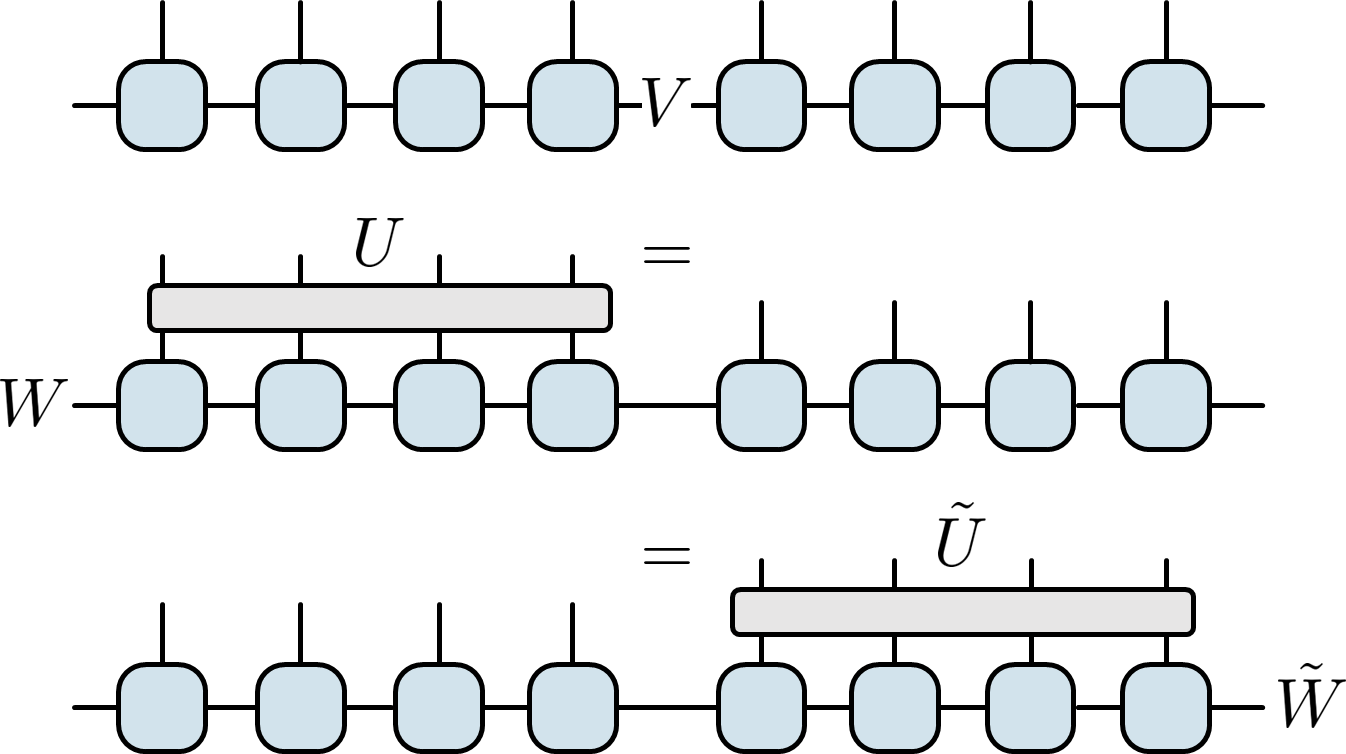}
    \caption{The generalized push-through relations considered in Appendix \ref{app:asymmetric-pushthrough}, pictured for $N=4$. The unitary operator $U$ ($\tilde{U}$) is used to push $V$ to the left (right) boundary of the MPS.}
    \label{fig:gen_pushthrough}
\end{figure}

Here, we remove some of the assumptions made in the main text about the structure of the push-through relations which allows more general measurement and feedforward protocols. We then show that we can derive the same constraint of a flat entanglement spectrum as in the main text. The relations that we assume are the following,
\begin{subequations} \label{eq:gen_pushthrough}
\begin{align}
    A^{i_1}\cdots A^{i_{N}}V = \sum_{j_1,\dots,j_{N}} U^{i_1,\dots, i_{N}}_{j_1,\dots,j_{N}} WA^{j_1} \cdots A^{j_{N}} 
    \, , \label{eq:gen_pushthrough_a} \\
    VA^{i_1}\cdots A^{i_{N}} = \sum_{j_1,\dots,j_{N}} \tilde{U}^{i_1,\dots, i_{N}}_{j_1,\dots,j_{N}} A^{j_1} \cdots A^{j_{N}} \tilde{W}
    \, , \label{eq:gen_pushthrough_b}
\end{align}
\end{subequations}
which hold for some complete set of byproduct operators $V$ and all $N$ where $U$, $W$, $\tilde{U}$, and $\tilde{W}$ (which all implicitly depend on $V$ and $N$) are assumed to be unitary since they are the correction operators we need to apply. We do not assume the byproduct operators $V$ are unitary, but we will later show that this must be true. 

These relations, shown in Fig.~\ref{fig:gen_pushthrough}, mean that there are some unitary operators that can be applied to push $V$ all the way to the left or right boundary and then annihilate it for all lengths $N$, thereby allowing us to prepare MPS of arbitrary length via fusion. Assuming that $V$ can be pushed to either the left or right has practical motivation, since it means we can always push $V$ to the nearest boundary, reducing the number of correction unitaries that we need to apply. The push-through relations in the main text \eqref{eq:mps_symm} satisfy this definition with $W=\tilde{W}=V$ and $U=\tilde{U}=u^{\otimes N}$. In the more general case of Eq.~\eqref{eq:gen_pushthrough}, $V$ is able change to another operator as it is pushed through the tensors, and the correction unitary $U$ need not be translationally invariant or even onsite. The former case can occur in the most general symmetry relation for non-normal MPS, where $W=V\Theta$ and $\Theta$ is a matrix of phases in each block \cite{Sanz2009}, while the latter case covers systems with spatially modulated symmetries \cite{Stephen2019,stephen2022universal,han2023topological} and allows for correction unitaries that are FDQCs, as are required in SIMPS fusion.

Now we rederive the constraint of flat entanglement spectrum with these generalized assumptions. The derivation has two steps: (i) proving the entanglement spectrum is flat in each block of the MPS and (ii) proving that all blocks have equal size.

\subsection{Flat entanglement spectrum in each block}
\label{app:flat-spectrum}

Observe that Eq.~\eqref{eq:gen_pushthrough_a} implies that the MPS transfer matrix satisfies
\begin{equation}
    (W\otimes \bar{W}) T^N  = T^N (V\otimes \bar{V})
    \, .
\end{equation}
Acting with $\bra{\Omega}$ on the left gives,
\begin{equation} \label{eq:tnvv}
    \bra{\Omega} T^N  = \bra{\Omega} T^N (V\otimes \bar{V})
    \, .
\end{equation}
where we used the unitarity of $W$. Now, we can use completeness of the operators $V$ to get,
\begin{equation}
    \bra{\Omega} T^N  = \bra{\Omega} T^N \ketbra{\Omega}{\Omega}
    \, .
\end{equation}
In the limit of large $N$, $T^N$ converges exponentially quickly to $\Pi$, the projector onto fixed points.
We therefore arrive at the constraint,
\begin{equation}
    \bra{\Omega} \Pi = \braket{\Omega | \Pi | \Omega}\bra{\Omega}
    \, .
\end{equation}
Acting with $\Pi$ on the right and using the fact that $\Pi^2=\Pi$ shows that $\braket{\Omega | \Pi | \Omega}=0$ or 1. Assume for now that the MPS is normal, so $N_b=1$. Then, the quantum Perron-Frobenius theorem ensures that the left and right fixed points are positive~\cite{Cirac2021} such that $\braket{\Omega | \Pi | \Omega}= \braket{\Omega|R}\braket{L|\Omega}\propto \Tr \sigma_R\Tr\sigma_L>0$, so $\braket{\Omega | \Pi | \Omega}=1$. If the MPS is not normal then we can make the same argument block-by-block. Namely, using Eq.~\eqref{eq:projector_decomp} and the fact that $|\Omega\rangle \propto \bigoplus_{\alpha,\beta} \delta_{\alpha\beta} \sqrt{D_\alpha} |\Omega_\alpha\rangle$, where $|\Omega_\alpha\rangle$ is shorthand for the normalized $D_\alpha$-dimensional qudit Bell state, we can decompose $\braket{\Omega | \Pi | \Omega}\propto \sum_\alpha D_\alpha \braket{\Omega_\alpha | \Pi_\alpha | \Omega_\alpha}$ (where the implied proportionality constant is positive). Since every block $\alpha$ corresponds to a normal MPS, we see that $\braket{\Omega_\alpha | \Pi_\alpha | \Omega_\alpha} > 0$ for all $\alpha$, so the sum is positive too and $\braket{\Omega | \Pi | \Omega} = 1$. Therefore $\bra{\Omega}$ must be a left fixed point of $T$. If we instead used Eq.~\eqref{eq:gen_pushthrough_b}, we would similarly find that $\ket{\Omega}$ is a right fixed-point of $T$. Using Eq.~\eqref{eq:tnvv} now gives $\bra{\Omega}=\bra{\Omega}V\otimes \bar{V}$, which implies that $V$ must be unitary, as claimed.

For normal MPS, where the fixed-points are unique, these results immediately imply a flat entanglement spectrum using Eq.~\eqref{eq:sigma-decomp}.
For non-normal MPS, using the fact that $\Pi$ has the same block structure \eqref{eq:projector_decomp}, it follows that $\ket{\Omega_\alpha}$ ($\bra{\Omega_\alpha}$) is a right (left) eigenvector of the block $\Pi_\alpha$. That is, the entanglement spectrum is flat in every block $\alpha$. However, to have the full spectrum be flat, we need all blocks to have equal size. This is what we prove in the next section.

\subsection{Equal block size}

Now we show that, in the case of non-normal MPS, fusion is only possible if the blocks $\alpha$ are all of the same size $D_\alpha = D/N_b$. For simplicity, we assume the injectivity length $L_0=1$ (or, alternatively, that we have blocked at least $L_0$ sites). Let $P_\alpha$ be the projector onto the states belonging to the block $\alpha$. Then, $P_\alpha A^i P_\beta = \delta_{\alpha\beta} A^i_\alpha$ follows from Eq.~\eqref{eqn:Ai-block-injective}. Now, Eq.~\eqref{eq:gen_pushthrough} with $N=1$ reads,
\begin{subequations} 
\begin{align}
    \sum_j U^i_jA^j &= WA^iV^{\dagger} \, , \\
    \sum_j \tilde{U}^i_jA^j &= V A^i \tilde{W}^{\dagger}
    \label{eqn:most-general-pushthrough}
\end{align}
\end{subequations}

Since the left-hand side of Eq.~\eqref{eqn:most-general-pushthrough} is a linear combination of the $A^i$, it must possess the same block structure as $A^i$. Inserting resolutions of identity $\sum_\alpha P_\alpha$ into~\eqref{eqn:most-general-pushthrough}, we deduce that,
\begin{equation}
    \sum_\beta (P_\alpha V P_\beta) A_\beta^i  (P_\beta W^\dagger P_\gamma) = 0 \quad \forall \alpha \neq \gamma.
    \label{eqn:off-diag-constraint}
\end{equation}
Using the shorthand notation $P_\alpha V P_\beta \equiv V_{\alpha\beta}$ (regarded as a $D_\alpha \times D_\beta$ matrix) and $P_\beta W^\dagger P_\gamma \equiv (W^\dagger)_{\beta\gamma} = (W_{\gamma\beta})^\dagger$,
we write Eq.~\eqref{eqn:off-diag-constraint} in an enlarged space 
\begin{equation}
    \bigoplus_\beta (V_{\alpha\beta} \otimes \bar{W}_{\gamma\beta}) \ket{A_\beta^i}  = 0 \quad \forall \alpha \neq \gamma.
\end{equation}
Block injectivity requires that this equation must hold for all vectors $\ket{A_\beta^i}$, and we find that the matrix $V_{\alpha\beta} \otimes \bar{W}_{\gamma\beta}$ must strictly vanish $\forall \beta,  \alpha \neq \gamma$. 
We therefore arrive at $V_{\alpha\beta} = 0$ or $\bar{W}_{\gamma\beta} = 0$. Given this structure, it is convenient to parameterize the block $V_{\alpha\beta}$ as $v_{\alpha\beta} \mathbb{V}_{\alpha\beta}$ (no sum), where $v_{\alpha\beta} \in \{ 0, 1 \}$ keeps track of the zero blocks and $\mathbb{V}_{\alpha\beta}$ the matrix part. Similarly, we write $W_{\alpha\beta} = w_{\alpha\beta} \mathbb{W}_{\alpha\beta}$.

Unitarity of $V$ (which was demonstrated in the previous section) implies that, for each $\alpha$,
\begin{equation}
    \sum_\beta V_{\alpha\beta} (V_{\alpha\beta})^\dagger = \sum_\beta (V_{\beta\alpha})^\dagger V_{\beta\alpha} = I_\alpha
    \, .
\end{equation}
Therefore, at least one $v_{\alpha\beta}$ must be non-zero in each row and in each column, else the sum vanishes. Suppose that $v_{\alpha\beta}$ has  \emph{exactly} one nonzero entry in every row and every column. Then $V_{\alpha\beta} = 0$ or $\bar{W}_{\gamma\beta} = 0$ $\forall \beta,  \alpha \neq \gamma$ uniquely fixes $w_{\alpha\beta} = v_{\alpha\beta}$. If one row of $v_{\alpha\beta}$ contains more than one non-zero entry then $w_{\alpha\beta}$ necessarily contains a row that vanishes, violating unitarity of $W$. We deduce that $v_{\alpha\beta}$ and $w_{\alpha\beta}$ must be identical permutation matrices. If $v_{\alpha\beta} = 1$ then we require,
\begin{equation}
    \mathbb{V}_{\alpha\beta} \mathbb{V}_{\alpha\beta}^\dagger = I_\alpha \quad \text{and} \quad \mathbb{V}_{\alpha\beta}^\dagger \mathbb{V}_{\alpha\beta} = I_\beta 
    \, .
\end{equation}
That is, $\mathbb{V}_{\alpha\beta}$ must be unitary and we must have $D_{\alpha} = D_{\beta}$. If $D_{\alpha} \neq D_{\beta}$ then $v_{\alpha\beta}$ must vanish and the $V$ cannot form a complete basis for $D \times D$ matrices. It follows that, if we require that the $V$ form a complete basis, then 
\begin{enumerate*}[label={(\roman*)}]
    \item all blocks must be the same size, $D_\alpha = D/N_b$
    \item the $V$ and $W$ take the form $V = P (\bigoplus_\alpha V_\alpha)$ and $W = P (\bigoplus_\alpha W_\alpha)$, where $P$ is a permutation matrix that permutes blocks and $V_\alpha$, $W_\alpha$ are unitary matrices.
\end{enumerate*}
We remark that the latter condition is consistent with the general push-through relations that arise from global symmetries in non-normal MPS \cite{Sanz2009}.

\section{Majumdar-Ghosh model} \label{sec:mg}

In this Appendix we discuss the precise connections between the non-normal MPS example from Eq.~\eqref{eqn:non-normal-state} and the Majumdar-Ghosh (MG) state~\cite{Majumdar1969}. The MG model is defined by the Hamiltonian
\begin{equation}
    H = 
    \sum_i \boldsymbol{\sigma}_{i} \cdot \boldsymbol{\sigma}_{i+1} + \frac12 \sum_i \boldsymbol{\sigma}_{i} \cdot \boldsymbol{\sigma}_{i+2}
    \, ,
    \label{eqn:MG-Hamiltonian}
\end{equation}
where $\boldsymbol{\sigma}_i = (X_i, Y_i, Z_i)$ is the vector of Pauli matrices. With PBC the Hamiltonian~\eqref{eqn:MG-Hamiltonian} has two ground states corresponding to the two distinct pairings of nearest neighbors into singlets. The symmetric superposition of these ground states is often called the MG state,
\begin{equation}
    \MG = \bigotimes_i |Y\rangle_{2i,2i+1} + \bigotimes_i |Y\rangle_{2i-1,2i}\ ,
\end{equation}
where $|Y\rangle = (iY\otimes I)|\Omega_2\rangle= (|01\rangle - |10\rangle)/\sqrt{2}$ is the singlet Bell state. 

We now show that $\MG$ is related to the state $|\phi\rangle$ \eqref{eqn:non-normal-state} via a local unitary change of basis. Specifically, we have,
\begin{equation}
    \prod_i H_{2i}Y_{2i} \MG = \bigotimes_i |H\rangle_{2i,2i+1} + (-1)^{N/2} \bigotimes_i |H\rangle_{2i-1,2i}
\end{equation}
where $|H\rangle = (H\otimes I)|\Omega_2\rangle$ and $H=(X+Z)/\sqrt{2}$ is the Hadamard matrix. Observe that $|H\rangle = \CZ \ket{++}$. Then, we have,
\begin{equation}
    \prod_i \CZ_{2i,2i+1} H_{2i}Y_{2i} \MG 
    = \ket{++\cdots +} + (-1)^{N/2}|C\rangle
\end{equation}
which matches Eq.~\eqref{eqn:non-normal-state} when $N/2$ is even. 

\section{Details of cocycles and anomalous SIMPS} \label{app:anom_calc}

Here we give details on the definitions and properties of 3-cocycles, following Ref.~\cite{Cheng2023}, and then derive some relations about the various operators defined in Sec.~\ref{sec:omega_simps}. A 3-cocycle of a group $G$ is a function $\omega:G\times G\times G\rightarrow U(1)$ satisfying the relation,
\begin{equation} \label{eq:cocycle_cond}
    \omega(g,h,k)\omega(g,hk,\ell)\omega(h,k,\ell) = \omega(gh,k,\ell)\omega(g,h,k\ell)
\end{equation}
Note that trilinear functions, as considered in the main text, are automatically cocycles. A special subset of 3-cocycles is given by the 3-coboundaries, which are functions of the form,
\begin{equation} \label{eq:coboundary}
    \omega(g,h,k)=\frac{\beta(h,k)\beta(g,hk)}{\beta(g,h)\beta(gh,k)}
\end{equation}
for any function $\beta:G\times G\rightarrow U(1)$. The set of 3-cocycles modulo 3-coboundaries forms an abelian group $H^3(G,U(1))$ whose elements are called cohomology classes. As discussed in the main text, any representation of a group $G$ using finite-depth quantum circuits (or more generally matrix product operators \cite{GarreRubio2023classifyingphases}) can be assigned a 3-cocycle, and the cohomology class this cocycle belongs to determines the anomaly of this symmetry.

Now we prove that $\prod_i L(h)_i$ and $\prod_i u^\omega(g)_{i,i+1}$, as defined in Sec.~\ref{sec:omega_simps}, commute for all $g,h\in G$.
First, note that the 3-cocycle condition Eq.~\ref{eq:cocycle_cond} can be rewritten in the following form,
\begin{equation}
\omega(g,hk,k^{-1}\ell)\omega(h,k,k^{-1}\ell)=\frac{\omega(gh,k,k^{-1}\ell)\omega(g,h,\ell)}{\omega(g,h,k)}
 .
\end{equation}
We can use this to derive,
\begin{equation} \label{eq:anom_1}
\begin{aligned}
    &[L(h)^\dagger \otimes L(h)^\dagger]u^\omega(g)[L(h)\otimes L(h)]\\
    &=u^\omega(gh)u^\omega(h)^{-1} [v^\omega(g,h)^{-1}\otimes v^\omega(g,h)] \\
    &=u^\omega(g) [v^\omega(g,h)^{-1}\otimes v^\omega(g,h)]
\end{aligned}
\end{equation}
where $v^\omega(g,h)|k\rangle =\omega(g,h,k)|k\rangle$, and we used the fact that $g\mapsto u^\omega(g)$ is a representation.
From this, we can straightforwardly see that $\prod_i L(h)_i$ and $\prod_i u^\omega(g)_{i,i+1}$ commute on periodic boundaries since the $v^\omega(g,h)$ and $v^\omega(g,h)^{-1}$ coming from neighboring gates in the product will cancel out.

We now provide the explicit form of the 3-site correction unitary $r^\omega(g)$ used in the fusion of anomalous SIMPS, as described in Sec.~\ref{sec:omega_simps}. In order to derive the form of $r^\omega(g)$, we invoke the circuit representation of the SIMPS tensor $B_\omega$ given in Eq.~\eqref{eq:omega_simps}, namely, using the definition $u^\omega(g)= \sum_{h,k\in G} \omega(g,h,h^{-1}k)\ket{h,k}\bra{ h,k }$ we may write
\begin{align}
    \ket{B_\omega} &\propto \sum_g u^\omega(g)_{12}\otimes \ket{g}\bra{g}_3 \sum_{h,k,\ell\in G} \ket{h,k,\ell,\ell} \notag  \\
    &= \sum_{h,k,\ell\in G} \omega(\ell, h, h^{-1}k)\ket{h,k,\ell,\ell} 
\end{align}
up to normalization. 
Now, the symmetry in Eq.~\eqref{eqn:omega-simps-symmetries} tells us that $L(g)^\dagger = \sum_{h\in G} \ket{g^{-1}h}\bra{h}$ and the diagonal 3-site unitary $r^\omega(g)$ act identically on the SIMPS tensor. Note that, according to our conventions, $L(g)^\dagger$ and $r^{\omega}(g)$ act on qubits $2$ and $\{ 1, 2, 4 \}$, respectively. It follows that
\begin{equation}
    L(g)_2^\dagger \ket{\psi_\omega} \propto 
    \sum_{h,k,\ell \in G}\omega(\ell, h, h^{-1} g k) \ket{h,k,\ell,\ell}
    \, .
\end{equation}
Therefore, the action of $L(g)^\dagger$ can be compensated by the action of the diagonal operator $r^\omega(g)$, which acts on states as
\begin{equation}
    r^\omega(g)  \ket{h,k,\ell} = \frac{\omega(\ell, h, h^{-1} g k)}{\omega(\ell, h, h^{-1} k)} \ket{h,k,\ell}
    \, .
\end{equation}

\bibliography{biblio}

@article{Else2014,
  title = {Classifying symmetry-protected topological phases through the anomalous action of the symmetry on the edge},
  author = {Else, Dominic V. and Nayak, Chetan},
  journal = {Phys. Rev. B},
  volume = {90},
  issue = {23},
  pages = {235137},
  numpages = {19},
  year = {2014},
  month = {Dec},
  publisher = {American Physical Society},
  doi = {10.1103/PhysRevB.90.235137},
  url = {https://link.aps.org/doi/10.1103/PhysRevB.90.235137}
}

@phdthesis{propitius1995topologicalinteractionsbrokengauge,
      title={Topological interactions in broken gauge theories}, 
      author={Mark de Wild Propitius},
      year={1995},
      eprint={hep-th/9511195},
      archivePrefix={arXiv},
      primaryClass={hep-th},
      url={https://arxiv.org/abs/hep-th/9511195}, 
      school={Institute for Theoretical Physics Amsterdam}
}

@article{GarreRubio2023classifyingphases,
  doi = {10.22331/q-2023-02-21-927},
  url = {https://doi.org/10.22331/q-2023-02-21-927},
  title = {Classifying phases protected by matrix product operator symmetries using matrix product states},
  author = {Garre-Rubio, Jos{\'{e}} and Lootens, Laurens and Moln{\'{a}}r, Andr{\'{a}}s},
  journal = {{Quantum}},
  issn = {2521-327X},
  publisher = {{Verein zur F{\"{o}}rderung des Open Access Publizierens in den Quantenwissenschaften}},
  volume = {7},
  pages = {927},
  month = feb,
  year = {2023}
}

@article{Malz2024,
  title = {Preparation of Matrix Product States with Log-Depth Quantum Circuits},
  author = {Malz, Daniel and Styliaris, Georgios and Wei, Zhi-Yuan and Cirac, J. Ignacio},
  journal = {Phys. Rev. Lett.},
  volume = {132},
  issue = {4},
  pages = {040404},
  numpages = {9},
  year = {2024},
  month = {Jan},
  publisher = {American Physical Society},
  doi = {10.1103/PhysRevLett.132.040404},
  url = {https://link.aps.org/doi/10.1103/PhysRevLett.132.040404}
}

@article{piroli2024approximating,
  title = {Approximating Many-Body Quantum States with Quantum Circuits and Measurements},
  author = {Piroli, Lorenzo and Styliaris, Georgios and Cirac, J. Ignacio},
  journal = {Phys. Rev. Lett.},
  volume = {133},
  issue = {23},
  pages = {230401},
  numpages = {7},
  year = {2024},
  month = {Dec},
  publisher = {American Physical Society},
  doi = {10.1103/PhysRevLett.133.230401},
  url = {https://link.aps.org/doi/10.1103/PhysRevLett.133.230401}
}

@article{gunn2023phases,
  title = {Phases of matrix product states with symmetric quantum circuits and symmetric measurements with feedforward},
  author = {Gunn, David and Styliaris, Georgios and Kraft, Tristan and Kraus, Barbara},
  journal = {Phys. Rev. B},
  volume = {111},
  issue = {11},
  pages = {115110},
  numpages = {26},
  year = {2025},
  month = {Mar},
  publisher = {American Physical Society},
  doi = {10.1103/PhysRevB.111.115110},
  url = {https://link.aps.org/doi/10.1103/PhysRevB.111.115110}
}

@article{OBrien2018,
  title = {Lattice Supersymmetry and Order-Disorder Coexistence in the Tricritical {I}sing Model},
  author = {O'Brien, Edward and Fendley, Paul},
  journal = {Phys. Rev. Lett.},
  volume = {120},
  issue = {20},
  pages = {206403},
  numpages = {5},
  year = {2018},
  month = {May},
  publisher = {American Physical Society},
  doi = {10.1103/PhysRevLett.120.206403},
  url = {https://link.aps.org/doi/10.1103/PhysRevLett.120.206403}
}

@article{Sanz2009,
  title = {Matrix product states: Symmetries and two-body Hamiltonians},
  author = {Sanz, M. and Wolf, M. M. and P\'erez-Garc\'{\i}a, D. and Cirac, J. I.},
  journal = {Phys. Rev. A},
  volume = {79},
  issue = {4},
  pages = {042308},
  numpages = {10},
  year = {2009},
  month = {Apr},
  publisher = {American Physical Society},
  doi = {10.1103/PhysRevA.79.042308},
  url = {https://link.aps.org/doi/10.1103/PhysRevA.79.042308}
}

@Article{Cheng2023,
	title={{Lieb-Schultz-Mattis, Luttinger, and 't Hooft - anomaly matching in lattice systems}},
	author={Meng Cheng and Nathan Seiberg},
	journal={SciPost Phys.},
	volume={15},
	pages={051},
	year={2023},
	publisher={SciPost},
	doi={10.21468/SciPostPhys.15.2.051},
	url={https://scipost.org/10.21468/SciPostPhys.15.2.051},
}

@article{Garratt2023,
  title = {Measurements Conspire Nonlocally to Restructure Critical Quantum States},
  author = {Garratt, Samuel J. and Weinstein, Zack and Altman, Ehud},
  journal = {Phys. Rev. X},
  volume = {13},
  issue = {2},
  pages = {021026},
  numpages = {24},
  year = {2023},
  month = {May},
  publisher = {American Physical Society},
  doi = {10.1103/PhysRevX.13.021026},
  url = {https://link.aps.org/doi/10.1103/PhysRevX.13.021026}
}

@article{Ippoliti2023,
  title = {Dynamical Purification and the Emergence of Quantum State Designs from the Projected Ensemble},
  author = {Ippoliti, Matteo and Ho, Wen Wei},
  journal = {PRX Quantum},
  volume = {4},
  issue = {3},
  pages = {030322},
  numpages = {28},
  year = {2023},
  month = {Aug},
  publisher = {American Physical Society},
  doi = {10.1103/PRXQuantum.4.030322},
  url = {https://link.aps.org/doi/10.1103/PRXQuantum.4.030322}
}

@article{Cotler2023,
  title = {Emergent Quantum State Designs from Individual Many-Body Wave Functions},
  author = {Cotler, Jordan S. and Mark, Daniel K. and Huang, Hsin-Yuan and Hern\'andez, Felipe and Choi, Joonhee and Shaw, Adam L. and Endres, Manuel and Choi, Soonwon},
  journal = {PRX Quantum},
  volume = {4},
  issue = {1},
  pages = {010311},
  numpages = {29},
  year = {2023},
  month = {Jan},
  publisher = {American Physical Society},
  doi = {10.1103/PRXQuantum.4.010311},
  url = {https://link.aps.org/doi/10.1103/PRXQuantum.4.010311}
}

@article{Lu2023mixed,
  title = {Mixed-State Long-Range Order and Criticality from Measurement and Feedback},
  author = {Lu, Tsung-Cheng and Zhang, Zhehao and Vijay, Sagar and Hsieh, Timothy H.},
  journal = {PRX Quantum},
  volume = {4},
  issue = {3},
  pages = {030318},
  numpages = {25},
  year = {2023},
  month = {Aug},
  publisher = {American Physical Society},
  doi = {10.1103/PRXQuantum.4.030318},
  url = {https://link.aps.org/doi/10.1103/PRXQuantum.4.030318}
}

@article{Lavasani2023,
  title = {Monitored quantum dynamics and the {K}itaev spin liquid},
  author = {Lavasani, Ali and Luo, Zhu-Xi and Vijay, Sagar},
  journal = {Phys. Rev. B},
  volume = {108},
  issue = {11},
  pages = {115135},
  numpages = {26},
  year = {2023},
  month = {Sep},
  publisher = {American Physical Society},
  doi = {10.1103/PhysRevB.108.115135},
  url = {https://link.aps.org/doi/10.1103/PhysRevB.108.115135}
}

@article{Tantivasadakarn2023hierarchy,
  title = {Hierarchy of Topological Order From Finite-Depth Unitaries, Measurement, and Feedforward},
  author = {Tantivasadakarn, Nathanan and Vishwanath, Ashvin and Verresen, Ruben},
  journal = {PRX Quantum},
  volume = {4},
  issue = {2},
  pages = {020339},
  numpages = {25},
  year = {2023},
  month = {Jun},
  publisher = {American Physical Society},
  doi = {10.1103/PRXQuantum.4.020339},
  url = {https://link.aps.org/doi/10.1103/PRXQuantum.4.020339}
}

@Article{Sahinoglu2021,
author={{\c{S}}ahino{\u{g}}lu, Mehmet Burak
and Williamson, Dominic
and Bultinck, Nick
and Mari{\"e}n, Micha{\"e}l
and Haegeman, Jutho
and Schuch, Norbert
and Verstraete, Frank},
title={Characterizing Topological Order with Matrix Product Operators},
journal={Annales Henri Poincar{\'e}},
year={2021},
month={Feb},
day={01},
volume={22},
number={2},
pages={563-592},
issn={1424-0661},
doi={10.1007/s00023-020-00992-4},
url={https://doi.org/10.1007/s00023-020-00992-4}
}

@misc{lee2022decoding,
      title={Decoding Measurement-Prepared Quantum Phases and Transitions: {F}rom {I}sing model to gauge theory, and beyond}, 
      author={Jong Yeon Lee and Wenjie Ji and Zhen Bi and Matthew P. A. Fisher},
      year={2022},
      eprint={2208.11699},
      archivePrefix={arXiv},
      primaryClass={cond-mat.str-el}
}

@Article{Choi2023,
author={Choi, Joonhee
and Shaw, Adam L.
and Madjarov, Ivaylo S.
and Xie, Xin
and Finkelstein, Ran
and Covey, Jacob P.
and Cotler, Jordan S.
and Mark, Daniel K.
and Huang, Hsin-Yuan
and Kale, Anant
and Pichler, Hannes
and Brand{\~a}o, Fernando G. S. L.
and Choi, Soonwon
and Endres, Manuel},
title={Preparing random states and benchmarking with many-body quantum chaos},
journal={Nature},
year={2023},
month={Jan},
day={01},
volume={613},
number={7944},
pages={468-473},
issn={1476-4687},
doi={10.1038/s41586-022-05442-1},
url={https://doi.org/10.1038/s41586-022-05442-1}
}

@Article{Hoke2023,
author={Hoke, J. C.
and others},
title={Measurement-induced entanglement and teleportation on a noisy quantum processor},
journal={Nature},
year={2023},
month={Oct},
day={01},
volume={622},
number={7983},
pages={481-486},
issn={1476-4687},
doi={10.1038/s41586-023-06505-7},
url={https://doi.org/10.1038/s41586-023-06505-7}
}

@article{DeCross2023,
  title = {Qubit-Reuse Compilation with Mid-Circuit Measurement and Reset},
  author = {DeCross, Matthew and Chertkov, Eli and Kohagen, Megan and Foss-Feig, Michael},
  journal = {Phys. Rev. X},
  volume = {13},
  issue = {4},
  pages = {041057},
  numpages = {22},
  year = {2023},
  month = {Dec},
  publisher = {American Physical Society},
  doi = {10.1103/PhysRevX.13.041057},
  url = {https://link.aps.org/doi/10.1103/PhysRevX.13.041057}
}

@article{bäumer2023efficient,
  title = {Efficient Long-Range Entanglement Using Dynamic Circuits},
  author = {B\"aumer, Elisa and Tripathi, Vinay and Wang, Derek S. and Rall, Patrick and Chen, Edward H. and Majumder, Swarnadeep and Seif, Alireza and Minev, Zlatko K.},
  journal = {PRX Quantum},
  volume = {5},
  issue = {3},
  pages = {030339},
  numpages = {20},
  year = {2024},
  month = {Aug},
  publisher = {American Physical Society},
  doi = {10.1103/PRXQuantum.5.030339},
  url = {https://link.aps.org/doi/10.1103/PRXQuantum.5.030339}
}

@article{Chen2011,
  title = {Two-dimensional symmetry-protected topological orders and their protected gapless edge excitations},
  author = {Chen, Xie and Liu, Zheng-Xin and Wen, Xiao-Gang},
  journal = {Phys. Rev. B},
  volume = {84},
  issue = {23},
  pages = {235141},
  numpages = {13},
  year = {2011},
  month = {Dec},
  publisher = {American Physical Society},
  doi = {10.1103/PhysRevB.84.235141},
  url = {https://link.aps.org/doi/10.1103/PhysRevB.84.235141}
}

@article{Kapustin2017,
  title = {Topological field theory and matrix product states},
  author = {Kapustin, Anton and Turzillo, Alex and You, Minyoung},
  journal = {Phys. Rev. B},
  volume = {96},
  issue = {7},
  pages = {075125},
  numpages = {13},
  year = {2017},
  month = {Aug},
  publisher = {American Physical Society},
  doi = {10.1103/PhysRevB.96.075125},
  url = {https://link.aps.org/doi/10.1103/PhysRevB.96.075125}
}

@article{Pezze2018,
  title = {Quantum metrology with nonclassical states of atomic ensembles},
  author = {Pezz\`e, Luca and Smerzi, Augusto and Oberthaler, Markus K. and Schmied, Roman and Treutlein, Philipp},
  journal = {Rev. Mod. Phys.},
  volume = {90},
  issue = {3},
  pages = {035005},
  numpages = {70},
  year = {2018},
  month = {Sep},
  publisher = {American Physical Society},
  doi = {10.1103/RevModPhys.90.035005},
  url = {https://link.aps.org/doi/10.1103/RevModPhys.90.035005}
}

@article{Wei2018,
author = {Tzu-Chieh Wei},
title = {Quantum spin models for measurement-based quantum computation},
journal = {Advances in Physics: X},
volume = {3},
number = {1},
pages = {1461026},
year = {2018},
publisher = {Taylor \& Francis},
doi = {10.1080/23746149.2018.1461026},
URL = {https://doi.org/10.1080/23746149.2018.1461026}
}

@article{Briegel1998,
  title = {Quantum Repeaters: The Role of Imperfect Local Operations in Quantum Communication},
  author = {Briegel, H.-J. and D\"ur, W. and Cirac, J. I. and Zoller, P.},
  journal = {Phys. Rev. Lett.},
  volume = {81},
  issue = {26},
  pages = {5932--5935},
  numpages = {0},
  year = {1998},
  month = {Dec},
  publisher = {American Physical Society},
  doi = {10.1103/PhysRevLett.81.5932},
  url = {https://link.aps.org/doi/10.1103/PhysRevLett.81.5932}
}

@article{Popp2005,
  title = {Localizable entanglement},
  author = {Popp, M. and Verstraete, F. and Mart\'{\i}n-Delgado, M. A. and Cirac, J. I.},
  journal = {Phys. Rev. A},
  volume = {71},
  issue = {4},
  pages = {042306},
  numpages = {18},
  year = {2005},
  month = {Apr},
  publisher = {American Physical Society},
  doi = {10.1103/PhysRevA.71.042306},
  url = {https://link.aps.org/doi/10.1103/PhysRevA.71.042306}
}

@Article{Acin2007,
author={Ac{\'i}n, Antonio
and Cirac, J. Ignacio
and Lewenstein, Maciej},
title={Entanglement percolation in quantum networks},
journal={Nature Physics},
year={2007},
month={Apr},
day={01},
volume={3},
number={4},
pages={256-259},
issn={1745-2481},
doi={10.1038/nphys549},
url={https://doi.org/10.1038/nphys549}
}

@article{Brennen2008,
  title = {Measurement-Based Quantum Computer in the Gapped Ground State of a Two-Body Hamiltonian},
  author = {Brennen, Gavin K. and Miyake, Akimasa},
  journal = {Phys. Rev. Lett.},
  volume = {101},
  issue = {1},
  pages = {010502},
  numpages = {4},
  year = {2008},
  month = {Jul},
  publisher = {American Physical Society},
  doi = {10.1103/PhysRevLett.101.010502},
  url = {https://link.aps.org/doi/10.1103/PhysRevLett.101.010502}
}

@article{Wei2012,
  title = {Two-dimensional {Affleck-Kennedy-Lieb-Tasaki} state on the honeycomb lattice is a universal resource for quantum computation},
  author = {Wei, Tzu-Chieh and Affleck, Ian and Raussendorf, Robert},
  journal = {Phys. Rev. A},
  volume = {86},
  issue = {3},
  pages = {032328},
  numpages = {19},
  year = {2012},
  month = {Sep},
  publisher = {American Physical Society},
  doi = {10.1103/PhysRevA.86.032328},
  url = {https://link.aps.org/doi/10.1103/PhysRevA.86.032328}
}

@article{Else2013,
  title = {Hidden symmetry-breaking picture of symmetry-protected topological order},
  author = {Else, Dominic V. and Bartlett, Stephen D. and Doherty, Andrew C.},
  journal = {Phys. Rev. B},
  volume = {88},
  issue = {8},
  pages = {085114},
  numpages = {10},
  year = {2013},
  month = {Aug},
  publisher = {American Physical Society},
  doi = {10.1103/PhysRevB.88.085114},
  url = {https://link.aps.org/doi/10.1103/PhysRevB.88.085114}
}

@article{Kennedy1992,
  title = {Hidden ${Z}_2 \times {Z}_2$ symmetry breaking in {H}aldane-gap antiferromagnets},
  author = {Kennedy, Tom and Tasaki, Hal},
  journal = {Phys. Rev. B},
  volume = {45},
  issue = {1},
  pages = {304--307},
  numpages = {0},
  year = {1992},
  month = {Jan},
  publisher = {American Physical Society},
  doi = {10.1103/PhysRevB.45.304},
  url = {https://link.aps.org/doi/10.1103/PhysRevB.45.304}
}

@article{mana2024kennedytasaki,
  title = {Kennedy-{T}asaki transformation and noninvertible symmetry in lattice models beyond one dimension},
  author = {Parayil Mana, Aswin and Li, Yabo and Sukeno, Hiroki and Wei, Tzu-Chieh},
  journal = {Phys. Rev. B},
  volume = {109},
  issue = {24},
  pages = {245129},
  numpages = {22},
  year = {2024},
  month = {Jun},
  publisher = {American Physical Society},
  doi = {10.1103/PhysRevB.109.245129},
  url = {https://link.aps.org/doi/10.1103/PhysRevB.109.245129}
}

@article{Majumdar1969,
    author = {Majumdar, Chanchal K. and Ghosh, Dipan K.},
    title = "{On Next‐Nearest‐Neighbor Interaction in Linear Chain. I}",
    journal = {Journal of Mathematical Physics},
    volume = {10},
    number = {8},
    pages = {1388-1398},
    year = {1969},
    month = {08},
    issn = {0022-2488},
    doi = {10.1063/1.1664978},
    url = {https://doi.org/10.1063/1.1664978}
}

@article{Zhou2003,
  title = {Quantum computation based on d-level cluster state},
  author = {Zhou, D. L. and Zeng, B. and Xu, Z. and Sun, C. P.},
  journal = {Phys. Rev. A},
  volume = {68},
  issue = {6},
  pages = {062303},
  numpages = {11},
  year = {2003},
  month = {Dec},
  publisher = {American Physical Society},
  doi = {10.1103/PhysRevA.68.062303},
  url = {https://link.aps.org/doi/10.1103/PhysRevA.68.062303}
}

@article{Stephen2019,
  doi = {10.22331/q-2019-05-20-142},
  url = {https://doi.org/10.22331/q-2019-05-20-142},
  title = {Subsystem symmetries, quantum cellular automata, and computational phases of quantum matter},
  author = {Stephen, David T. and Nautrup, Hendrik Poulsen and Bermejo-Vega, Juani and Eisert, Jens and Raussendorf, Robert},
  journal = {{Quantum}},
  issn = {2521-327X},
  publisher = {{Verein zur F{\"{o}}rderung des Open Access Publizierens in den Quantenwissenschaften}},
  volume = {3},
  pages = {142},
  month = may,
  year = {2019}
}

@article{stephen2022universal,
  title = {Universal Measurement-Based Quantum Computation in a One-Dimensional Architecture Enabled by Dual-Unitary Circuits},
  author = {Stephen, David T. and Ho, Wen Wei and Wei, Tzu-Chieh and Raussendorf, Robert and Verresen, Ruben},
  journal = {Phys. Rev. Lett.},
  volume = {132},
  issue = {25},
  pages = {250601},
  numpages = {7},
  year = {2024},
  month = {Jun},
  publisher = {American Physical Society},
  doi = {10.1103/PhysRevLett.132.250601},
  url = {https://link.aps.org/doi/10.1103/PhysRevLett.132.250601}
}

@article{han2023topological,
  title = {Topological quantum chains protected by dipolar and other modulated symmetries},
  author = {Han, Jung Hoon and Lake, Ethan and Lam, Ho Tat and Verresen, Ruben and You, Yizhi},
  journal = {Phys. Rev. B},
  volume = {109},
  issue = {12},
  pages = {125121},
  numpages = {19},
  year = {2024},
  month = {Mar},
  publisher = {American Physical Society},
  doi = {10.1103/PhysRevB.109.125121},
  url = {https://link.aps.org/doi/10.1103/PhysRevB.109.125121}
}

@article{Zhu2023,
  title = {Nishimori's Cat: Stable Long-Range Entanglement from Finite-Depth Unitaries and Weak Measurements},
  author = {Zhu, Guo-Yi and Tantivasadakarn, Nathanan and Vishwanath, Ashvin and Trebst, Simon and Verresen, Ruben},
  journal = {Phys. Rev. Lett.},
  volume = {131},
  issue = {20},
  pages = {200201},
  numpages = {9},
  year = {2023},
  month = {Nov},
  publisher = {American Physical Society},
  doi = {10.1103/PhysRevLett.131.200201},
  url = {https://link.aps.org/doi/10.1103/PhysRevLett.131.200201}
}

@article{Cirac2021,
  title = {Matrix product states and projected entangled pair states: Concepts, symmetries, theorems},
  author = {Cirac, J. Ignacio and P\'erez-Garc\'{\i}a, David and Schuch, Norbert and Verstraete, Frank},
  journal = {Rev. Mod. Phys.},
  volume = {93},
  issue = {4},
  pages = {045003},
  numpages = {65},
  year = {2021},
  month = {Dec},
  publisher = {American Physical Society},
  doi = {10.1103/RevModPhys.93.045003},
  url = {https://link.aps.org/doi/10.1103/RevModPhys.93.045003}
}

@article{Cirac2011,
  title = {Entanglement spectrum and boundary theories with projected entangled-pair states},
  author = {Cirac, J. Ignacio and Poilblanc, Didier and Schuch, Norbert and Verstraete, Frank},
  journal = {Phys. Rev. B},
  volume = {83},
  issue = {24},
  pages = {245134},
  numpages = {12},
  year = {2011},
  month = {Jun},
  publisher = {American Physical Society},
  doi = {10.1103/PhysRevB.83.245134},
  url = {https://link.aps.org/doi/10.1103/PhysRevB.83.245134}
}

@misc{friedman2023locality,
      title={Locality and error correction in quantum dynamics with measurement}, 
      author={Aaron J. Friedman and Chao Yin and Yifan Hong and Andrew Lucas},
      year={2023},
      eprint={2206.09929},
      archivePrefix={arXiv},
      primaryClass={quant-ph}
}

@article{PerezGarcia2006,
 author = {Perez-Garcia, D. and Verstraete, F. and Wolf, M. M. and Cirac, J. I.},
 title = {Matrix Product State Representations},
 journal = {Quantum Info. Comput.},
 issue_date = {July 2007},
 volume = {7},
 number = {5},
 month = {Jul},
 year = {2007},
 pages = {401--430},
 numpages = {30},
 optdoi = {10.26421/QIC7.5-6},
 url = {http://dl.acm.org/citation.cfm?id=2011832.2011833},
 acmid = {2011833},
 publisher = {Rinton Press, Incorporated},
 address = {Paramus, NJ},
}

@article{Raussendorf2001,
  title = {Persistent Entanglement in Arrays of Interacting Particles},
  author = {Briegel, Hans J. and Raussendorf, Robert},
  journal = {Phys. Rev. Lett.},
  volume = {86},
  issue = {5},
  pages = {910--913},
  numpages = {0},
  year = {2001},
  month = {Jan},
  publisher = {American Physical Society},
  doi = {10.1103/PhysRevLett.86.910},
  url = {https://link.aps.org/doi/10.1103/PhysRevLett.86.910}
}

@misc{bravyi2022adaptive,
      title={Adaptive constant-depth circuits for manipulating non-abelian anyons}, 
      author={Sergey Bravyi and Isaac Kim and Alexander Kliesch and Robert Koenig},
      year={2022},
      eprint={2205.01933},
      archivePrefix={arXiv},
      primaryClass={quant-ph}
}

@article{Piroli2021,
  title = {Quantum Circuits Assisted by Local Operations and Classical Communication: Transformations and Phases of Matter},
  author = {Piroli, Lorenzo and Styliaris, Georgios and Cirac, J. Ignacio},
  journal = {Phys. Rev. Lett.},
  volume = {127},
  issue = {22},
  pages = {220503},
  numpages = {6},
  year = {2021},
  month = {Nov},
  publisher = {American Physical Society},
  doi = {10.1103/PhysRevLett.127.220503},
  url = {https://link.aps.org/doi/10.1103/PhysRevLett.127.220503}
}

@article{Affleck1987,
  title = {Rigorous results on valence-bond ground states in antiferromagnets},
  author = {Affleck, Ian and Kennedy, Tom and Lieb, Elliott H. and Tasaki, Hal},
  journal = {Phys. Rev. Lett.},
  volume = {59},
  issue = {7},
  pages = {799--802},
  numpages = {0},
  year = {1987},
  month = {Aug},
  publisher = {American Physical Society},
  doi = {10.1103/PhysRevLett.59.799},
  opturl = {https://link.aps.org/doi/10.1103/PhysRevLett.59.799}
}

@article{PerezGarcia2008,
  title = {String Order and Symmetries in Quantum Spin Lattices},
  author = {P\'erez-Garc\'{\i}a, D. and Wolf, M. M. and Sanz, M. and Verstraete, F. and Cirac, J. I.},
  journal = {Phys. Rev. Lett.},
  volume = {100},
  issue = {16},
  pages = {167202},
  numpages = {4},
  year = {2008},
  month = {Apr},
  publisher = {American Physical Society},
  doi = {10.1103/PhysRevLett.100.167202},
  url = {https://link.aps.org/doi/10.1103/PhysRevLett.100.167202}
}

@article{Herringer2023classificationof,
  doi = {10.22331/q-2023-06-12-1041},
  url = {https://doi.org/10.22331/q-2023-06-12-1041},
  title = {Classification of measurement-based quantum wire in stabilizer {PEPS}},
  author = {Herringer, Paul and Raussendorf, Robert},
  journal = {{Quantum}},
  issn = {2521-327X},
  publisher = {{Verein zur F{\"{o}}rderung des Open Access Publizierens in den Quantenwissenschaften}},
  volume = {7},
  pages = {1041},
  month = jun,
  year = {2023}
}

@article{SIMPS,
  doi = {10.22331/q-2025-05-12-1738},
  url = {https://doi.org/10.22331/q-2025-05-12-1738},
  title = {Non-onsite symmetries and quantum teleportation in split-index matrix product states},
  author = {Stephen, David T.},
  journal = {{Quantum}},
  issn = {2521-327X},
  publisher = {{Verein zur F{\"{o}}rderung des Open Access Publizierens in den Quantenwissenschaften}},
  volume = {9},
  pages = {1738},
  month = may,
  year = {2025}
}

@article{Smith_AKLT,
  title = {Deterministic Constant-Depth Preparation of the {AKLT} State on a Quantum Processor Using Fusion Measurements},
  author = {Smith, Kevin C. and Crane, Eleanor and Wiebe, Nathan and Girvin, S.M.},
  journal = {PRX Quantum},
  volume = {4},
  issue = {2},
  pages = {020315},
  numpages = {24},
  year = {2023},
  month = {Apr},
  publisher = {American Physical Society},
  doi = {10.1103/PRXQuantum.4.020315},
  url = {https://link.aps.org/doi/10.1103/PRXQuantum.4.020315}
}

@article{tantivasadakarn2022longrange,
  title = {Long-Range Entanglement from Measuring Symmetry-Protected Topological Phases},
  author = {Tantivasadakarn, Nathanan and Thorngren, Ryan and Vishwanath, Ashvin and Verresen, Ruben},
  journal = {Phys. Rev. X},
  volume = {14},
  issue = {2},
  pages = {021040},
  numpages = {22},
  year = {2024},
  month = {Jun},
  publisher = {American Physical Society},
  doi = {10.1103/PhysRevX.14.021040},
  url = {https://link.aps.org/doi/10.1103/PhysRevX.14.021040}
}

@misc{verresen2022efficiently,
      title={Efficiently preparing {S}chr\"odinger's cat, fractons and non-{A}belian topological order in quantum devices}, 
      author={Ruben Verresen and Nathanan Tantivasadakarn and Ashvin Vishwanath},
      year={2022},
      eprint={2112.03061},
      archivePrefix={arXiv},
      primaryClass={quant-ph}
}

@article{Tantivasadakarn2023shortest,
  title = {Shortest Route to Non-{A}belian Topological Order on a Quantum Processor},
  author = {Tantivasadakarn, Nathanan and Verresen, Ruben and Vishwanath, Ashvin},
  journal = {Phys. Rev. Lett.},
  volume = {131},
  issue = {6},
  pages = {060405},
  numpages = {5},
  year = {2023},
  month = {Aug},
  publisher = {American Physical Society},
  doi = {10.1103/PhysRevLett.131.060405},
  url = {https://link.aps.org/doi/10.1103/PhysRevLett.131.060405}
}

@article{Lu2022Measurement,
  title = {Measurement as a Shortcut to Long-Range Entangled Quantum Matter},
  author = {Lu, Tsung-Cheng and Lessa, Leonardo A. and Kim, Isaac H. and Hsieh, Timothy H.},
  journal = {PRX Quantum},
  volume = {3},
  issue = {4},
  pages = {040337},
  numpages = {22},
  year = {2022},
  month = {Dec},
  publisher = {American Physical Society},
  doi = {10.1103/PRXQuantum.3.040337},
  url = {https://link.aps.org/doi/10.1103/PRXQuantum.3.040337}
}

@article{iqbal2023topological,
  title={Topological order from measurements and feed-forward on a trapped ion quantum computer},
  author={Iqbal, Mohsin and Tantivasadakarn, Nathanan and Gatterman, Thomas M and Gerber, Justin A and Gilmore, Kevin and Gresh, Dan and Hankin, Aaron and Hewitt, Nathan and Horst, Chandler V and Matheny, Mitchell and others},
  journal={Communications Physics},
  volume={7},
  number={1},
  pages={205},
  year={2024},
  publisher={Nature Publishing Group UK London},
  doi={10.1038/s42005-024-01698-3}
}

@misc{fossfeig2023experimental,
      title={Experimental demonstration of the advantage of adaptive quantum circuits}, 
      author={Michael Foss-Feig and Arkin Tikku and Tsung-Cheng Lu and Karl Mayer and Mohsin Iqbal and Thomas M. Gatterman and Justin A. Gerber and Kevin Gilmore and Dan Gresh and Aaron Hankin and Nathan Hewitt and Chandler V. Horst and Mitchell Matheny and Tanner Mengle and Brian Neyenhuis and Henrik Dreyer and David Hayes and Timothy H. Hsieh and Isaac H. Kim},
      year={2023},
      eprint={2302.03029},
      archivePrefix={arXiv},
      primaryClass={quant-ph}
}

@article{hart2024playing,
  title = {Playing Nonlocal Games across a Topological Phase Transition on a Quantum Computer},
  author = {Hart, Oliver and Stephen, David T. and Williamson, Dominic J. and Foss-Feig, Michael and Nandkishore, Rahul},
  journal = {Phys. Rev. Lett.},
  volume = {134},
  issue = {13},
  pages = {130602},
  numpages = {6},
  year = {2025},
  month = {Mar},
  publisher = {American Physical Society},
  doi = {10.1103/PhysRevLett.134.130602},
  url = {https://link.aps.org/doi/10.1103/PhysRevLett.134.130602}
}

@article{AKLT1987,
  title = {Rigorous results on valence-bond ground states in antiferromagnets},
  author = {Affleck, Ian and Kennedy, Tom and Lieb, Elliott H. and Tasaki, Hal},
  journal = {Phys. Rev. Lett.},
  volume = {59},
  issue = {7},
  pages = {799--802},
  numpages = {0},
  year = {1987},
  month = {Aug},
  publisher = {American Physical Society},
  doi = {10.1103/PhysRevLett.59.799},
  url = {https://link.aps.org/doi/10.1103/PhysRevLett.59.799}
}

@article{AKLT1988,
  title={Valence bond ground states in isotropic quantum antiferromagnets},
  author={Affleck, Ian and Kennedy, Tom and Lieb, Elliott H and Tasaki, Hal},
  journal={Communications in Mathematical Physics},
  volume={115},
  number={3},
  pages={477--528},
  year={1988},
  publisher={Springer},
  doi={10.1007/BF01218021}
}

@article{Bravyi2006,
  title = {Lieb-{R}obinson Bounds and the Generation of Correlations and Topological Quantum Order},
  author = {Bravyi, S. and Hastings, M. B. and Verstraete, F.},
  journal = {Phys. Rev. Lett.},
  volume = {97},
  issue = {5},
  pages = {050401},
  numpages = {4},
  year = {2006},
  month = {Jul},
  publisher = {American Physical Society},
  doi = {10.1103/PhysRevLett.97.050401},
  url = {https://link.aps.org/doi/10.1103/PhysRevLett.97.050401}
}

@article{Chen2010,
  title = {Local unitary transformation, long-range quantum entanglement, wave function renormalization, and topological order},
  author = {Chen, Xie and Gu, Zheng-Cheng and Wen, Xiao-Gang},
  journal = {Phys. Rev. B},
  volume = {82},
  issue = {15},
  pages = {155138},
  numpages = {28},
  year = {2010},
  month = {Oct},
  publisher = {American Physical Society},
  doi = {10.1103/PhysRevB.82.155138},
  url = {https://link.aps.org/doi/10.1103/PhysRevB.82.155138}
}

@article{Wahl2012,
  title = {Matrix product states with long-range localizable entanglement},
  author = {Wahl, T. B. and P\'erez-Garc\'{\i}a, D. and Cirac, J. I.},
  journal = {Phys. Rev. A},
  volume = {86},
  issue = {6},
  pages = {062314},
  numpages = {8},
  year = {2012},
  month = {Dec},
  publisher = {American Physical Society},
  doi = {10.1103/PhysRevA.86.062314},
  url = {https://link.aps.org/doi/10.1103/PhysRevA.86.062314}
}

@article{Schon2005,
  title = {Sequential Generation of Entangled Multiqubit States},
  author = {Sch\"on, C. and Solano, E. and Verstraete, F. and Cirac, J. I. and Wolf, M. M.},
  journal = {Phys. Rev. Lett.},
  volume = {95},
  issue = {11},
  pages = {110503},
  numpages = {4},
  year = {2005},
  month = {Sep},
  publisher = {American Physical Society},
  doi = {10.1103/PhysRevLett.95.110503},
  url = {https://link.aps.org/doi/10.1103/PhysRevLett.95.110503}
}

@article{Huang2015,
  title = {Quantum circuit complexity of one-dimensional topological phases},
  author = {Huang, Yichen and Chen, Xie},
  journal = {Phys. Rev. B},
  volume = {91},
  issue = {19},
  pages = {195143},
  numpages = {10},
  year = {2015},
  month = {May},
  publisher = {American Physical Society},
  doi = {10.1103/PhysRevB.91.195143},
  url = {https://link.aps.org/doi/10.1103/PhysRevB.91.195143}
}

@article{Pollmann2012,
  title = {Symmetry protection of topological phases in one-dimensional quantum spin systems},
  author = {Pollmann, Frank and Berg, Erez and Turner, Ari M. and Oshikawa, Masaki},
  journal = {Phys. Rev. B},
  volume = {85},
  issue = {7},
  pages = {075125},
  numpages = {9},
  year = {2012},
  month = {Feb},
  publisher = {American Physical Society},
  doi = {10.1103/PhysRevB.85.075125},
  url = {https://link.aps.org/doi/10.1103/PhysRevB.85.075125}
}

@article{Pollmann2010,
  title = {Entanglement spectrum of a topological phase in one dimension},
  author = {Pollmann, Frank and Turner, Ari M. and Berg, Erez and Oshikawa, Masaki},
  journal = {Phys. Rev. B},
  volume = {81},
  issue = {6},
  pages = {064439},
  numpages = {10},
  year = {2010},
  month = {Feb},
  publisher = {American Physical Society},
  doi = {10.1103/PhysRevB.81.064439},
  url = {https://link.aps.org/doi/10.1103/PhysRevB.81.064439}
}

@article{AnthonyLR2023,
doi = {10.1088/1361-6633/acfaae},
url = {https://dx.doi.org/10.1088/1361-6633/acfaae},
year = {2023},
month = {sep},
publisher = {IOP Publishing},
volume = {86},
number = {11},
pages = {116001},
author = {Chi-Fang (Anthony) Chen and Andrew Lucas and Chao Yin},
title = {Speed limits and locality in many-body quantum dynamics},
journal = {Reports on Progress in Physics}
}

@Article{seiberg2024noninvertible,
	title={{Non-invertible symmetries and {LSM}-type constraints on a tensor product {H}ilbert space}},
	author={Nathan Seiberg and Sahand Seifnashri and Shu-Heng Shao},
	journal={SciPost Phys.},
	volume={16},
	pages={154},
	year={2024},
	publisher={SciPost},
	doi={10.21468/SciPostPhys.16.6.154},
	url={https://scipost.org/10.21468/SciPostPhys.16.6.154},
}

@article{pino2021demonstration,
  title={Demonstration of the trapped-ion quantum CCD computer architecture},
  author={Pino, Juan M and Dreiling, Jennifer M and Figgatt, Caroline and Gaebler, John P and Moses, Steven A and Allman, MS and Baldwin, CH and Foss-Feig, Michael and Hayes, D and Mayer, K and others},
  journal={Nature},
  volume={592},
  number={7853},
  pages={209--213},
  year={2021},
  publisher={Nature Publishing Group UK London},
  doi={10.1038/s41586-021-03318-4}
}

@article{RyanAndersonQEC2021,
  title = {Realization of Real-Time Fault-Tolerant Quantum Error Correction},
  author = {Ryan-Anderson, C. and Bohnet, J. G. and Lee, K. and Gresh, D. and Hankin, A. and Gaebler, J. P. and Francois, D. and Chernoguzov, A. and Lucchetti, D. and Brown, N. C. and Gatterman, T. M. and Halit, S. K. and Gilmore, K. and Gerber, J. A. and Neyenhuis, B. and Hayes, D. and Stutz, R. P.},
  journal = {Phys. Rev. X},
  volume = {11},
  issue = {4},
  pages = {041058},
  numpages = {29},
  year = {2021},
  month = {Dec},
  publisher = {American Physical Society},
  doi = {10.1103/PhysRevX.11.041058},
  url = {https://link.aps.org/doi/10.1103/PhysRevX.11.041058}
}

@article{Corcoles2021Dynamic,
  title = {Exploiting Dynamic Quantum Circuits in a Quantum Algorithm with Superconducting Qubits},
  author = {C\'orcoles, A. D. and Takita, Maika and Inoue, Ken and Lekuch, Scott and Minev, Zlatko K. and Chow, Jerry M. and Gambetta, Jay M.},
  journal = {Phys. Rev. Lett.},
  volume = {127},
  issue = {10},
  pages = {100501},
  numpages = {6},
  year = {2021},
  month = {Aug},
  publisher = {American Physical Society},
  doi = {10.1103/PhysRevLett.127.100501},
  url = {https://link.aps.org/doi/10.1103/PhysRevLett.127.100501}
}

@article{RaussendorfOneWay,
  title = {A One-Way Quantum Computer},
  author = {Raussendorf, Robert and Briegel, Hans J.},
  journal = {Phys. Rev. Lett.},
  volume = {86},
  issue = {22},
  pages = {5188--5191},
  numpages = {0},
  year = {2001},
  month = {May},
  publisher = {American Physical Society},
  doi = {10.1103/PhysRevLett.86.5188},
  url = {https://link.aps.org/doi/10.1103/PhysRevLett.86.5188}
}

@article{RaussendorfLongRange,
  title = {Long-range quantum entanglement in noisy cluster states},
  author = {Raussendorf, Robert and Bravyi, Sergey and Harrington, Jim},
  journal = {Phys. Rev. A},
  volume = {71},
  issue = {6},
  pages = {062313},
  numpages = {6},
  year = {2005},
  month = {Jun},
  publisher = {American Physical Society},
  doi = {10.1103/PhysRevA.71.062313},
  url = {https://link.aps.org/doi/10.1103/PhysRevA.71.062313}
}

@article{Lin2021,
  title = {Real- and Imaginary-Time Evolution with Compressed Quantum Circuits},
  author = {Lin, Sheng-Hsuan and Dilip, Rohit and Green, Andrew G. and Smith, Adam and Pollmann, Frank},
  journal = {PRX Quantum},
  volume = {2},
  issue = {1},
  pages = {010342},
  numpages = {15},
  year = {2021},
  month = {Mar},
  publisher = {American Physical Society},
  doi = {10.1103/PRXQuantum.2.010342},
  url = {https://link.aps.org/doi/10.1103/PRXQuantum.2.010342}
}

@misc{RahulGeneralFusion1,
      title={Finite-Depth Preparation of Tensor Network States from Measurement}, 
      author={Rahul Sahay and Ruben Verresen},
      year={2024},
      eprint={2404.17087},
      archivePrefix={arXiv},
      primaryClass={quant-ph}
}

@article{RahulGeneralFusion2,
  title = {Classifying One-Dimensional Quantum States Prepared by a Single Round of Measurements},
  author = {Sahay, Rahul and Verresen, Ruben},
  journal = {PRX Quantum},
  volume = {6},
  issue = {1},
  pages = {010329},
  numpages = {42},
  year = {2025},
  month = {Feb},
  publisher = {American Physical Society},
  doi = {10.1103/PRXQuantum.6.010329},
  url = {https://link.aps.org/doi/10.1103/PRXQuantum.6.010329}
}

@article{SmithGeneralFusion,
  title = {Constant-Depth Preparation of Matrix Product States with Adaptive Quantum Circuits},
  author = {Smith, Kevin C. and Khan, Abid and Clark, Bryan K. and Girvin, S.M. and Wei, Tzu-Chieh},
  journal = {PRX Quantum},
  volume = {5},
  issue = {3},
  pages = {030344},
  numpages = {36},
  year = {2024},
  month = {Sep},
  publisher = {American Physical Society},
  doi = {10.1103/PRXQuantum.5.030344},
  url = {https://link.aps.org/doi/10.1103/PRXQuantum.5.030344}
}

\end{document}